\documentclass[journal]{IEEEtran}

\usepackage{mathtools,subfigure,amsfonts,makecell,changes,amssymb,multirow,booktabs,array,colortbl,pifont}
\usepackage[10pt]{moresize}

\newcommand{\cmark}{\ding{51}}%
\newcommand{\xmark}{\ding{55}}%

\definecolor{mygray}{gray}{.93}

\newlength{\Oldarrayrulewidth}
\newcommand{\Cline}[2]{%
	\noalign{\global\setlength{\Oldarrayrulewidth}{\arrayrulewidth}}%
	\noalign{\global\setlength{\arrayrulewidth}{#1}}\cline{#2}%
	\noalign{\global\setlength{\arrayrulewidth}{\Oldarrayrulewidth}}}

\ifCLASSINFOpdf
\else
\fi

\begin{document}
\title{Audio-Visual Speech Separation and Dereverberation with a Two-Stage Multimodal Network}

\author{Ke~Tan, Yong~Xu, Shi-Xiong~Zhang, Meng~Yu, and~Dong~Yu,~\IEEEmembership{Fellow,~IEEE}% <-this % stops a space
	\thanks{K. Tan is with the Department
		of Computer Science and Engineering, The Ohio State University, Columbus,
		OH, 43210-1277 USA (e-mail: tan.650@osu.edu). This work was performed when K. Tan was an intern at Tencent AI Lab, Bellevue, WA, USA.}% <-this % stops a space
	\thanks{Y. Xu, S.-X. Zhang, M. Yu and D. Yu are with Tencent AI Lab, Bellevue, WA, USA (e-mails: lucayongxu@tencent.com, auszhang@tencent.com, raymondmyu@tencent.com, dyu@tencent.com).}% <-this % stops a space
	}


\maketitle

\begin{abstract}
Background noise, interfering speech and room reverberation frequently distort target speech in real listening environments. In this study, we address joint speech separation and dereverberation, which aims to separate target speech from background noise, interfering speech and room reverberation. In order to tackle this fundamentally difficult problem, we propose a novel multimodal network that exploits both audio and visual signals. The proposed network architecture adopts a two-stage strategy, where a separation module is employed to attenuate background noise and interfering speech in the first stage and a dereverberation module to suppress room reverberation in the second stage. The two modules are first trained separately, and then integrated for joint training, which is based on a new multi-objective loss function. Our experimental results show that the proposed multimodal network yields consistently better objective intelligibility and perceptual quality than several one-stage and two-stage baselines. We find that our network achieves a 21.10\% improvement in ESTOI and a 0.79 improvement in PESQ over the unprocessed mixtures. Moreover, our network architecture does not require the knowledge of the number of speakers.
\end{abstract}


% \fi
\IEEEpeerreviewmaketitle

\section{Introduction}
\IEEEPARstart{I}{N AN} acoustic environment like a cocktail party, the human auditory system is remarkably capable of following a single target speech source in the presence of interfering speakers, background noise and room reverberation. Speech separation, also commonly known as the cocktail party problem, is the task of separating target speech from background interference~\cite{cherry1953some},~\cite{wang2018supervised}. Both interfering sounds from other sources and reverberation from surface reflections corrupt target speech, which can severely degrade speech intelligibility for human listeners, as well as the performance of computing systems for speech processing. Numerous research efforts have been made to improve the performance of speech separation for decades. Inspired by the concept of time-frequency (T-F) masking in computational auditory scene analysis (CASA), speech separation has been recently formulated as supervised learning, where discriminative patterns within target speech or background interference are learned from training data~\cite{wang2006computational}. Thanks to the use of deep learning, the performance of supervised speech separation has been substantially elevated in the last decade~\cite{wang2013towards},~\cite{wang2018supervised}. Producing high-quality separated speech in adverse acoustic environments, however, still remains a challenging problem.

Speaker separation has attracted considerable research attention in the last several years, of which the goal is to extract multiple speech sources, one for each speaker. Speaker-independent speech separation, where none of the speakers are required to be the same between training and testing, is susceptible to the label ambiguity (or permutation) problem~\cite{weng2015deep},~\cite{hershey2016deep}. Notable approaches to speaker-independent speech separation include deep clustering~\cite{hershey2016deep} and permutation-invariant training (PIT)~\cite{yu2017permutation}, which address the label ambiguity from different angles. Deep clustering treats speaker separation as spectral clustering, while PIT uses a dynamically calculated loss function for training. Many recent studies have extended these two approaches. For example, a dilated convolutional neural network (CNN) named TasNet is employed to perform time-domain speech separation in~\cite{luo2019conv}, where utterance-level PIT~\cite{kolbaek2017multitalker} is applied during training. An alternative way to resolve the label ambiguity is to use speaker-discriminative acoustic cues of the target speaker as an auxiliary input for separation. In a recent study~\cite{wang2018deep}, a pre-recorded short utterance from the target speaker is used as an anchor for attentional control, which selects the target speaker to be separated. Analogously, a speaker-discriminative embedding is produced by a speaker recognition network from a reference signal of the target speaker in~\cite{wang2019voicefilter}. The embedding vector, along with the spectrogram of the noisy mixture, is then fed into the separation network. A potential advantage of such approaches is that the knowledge of the number of speakers is not required.

Visual cues such as facial movements or lip movements of a speaker can supplement the information from the speaker's voice and thus facilitate speech perception, particularly in noisy environments~\cite{macleod1987quantifying},~\cite{massaro2014speech},~\cite{rosenblum2008speech}. Motivated by this finding, various algorithms have been developed to combine audio and visual signals to perform speech separation in a multimodal manner~\cite{rivet2014audiovisual}. There is recent interest in using deep neural networks (DNNs) to achieve this goal. Hou {\it et al.}~\cite{hou2018audio} designed an audio-visual speech enhancement framework based on multi-task learning. Their experimental results show that the audio-visual enhancement framework consistently outperforms the same architecture without visual inputs. A similar model is developed in~\cite{gabbay2018visual}, where a CNN is trained to directly estimate the magnitude spectrogram of clean speech from noisy speech and the input video. Moreover, Gabbay {\it et al.}~\cite{gabbay2018seeing} employs a video-to-speech method to synthesize speech, which is subsequently used to construct T-F masks for speech separation. Other related studies include~\cite{hou2016audio},~\cite{wu2016multi},~\cite{khan2018using}. 

Although the aforementioned deep learning based audio-visual approaches considerably elevate the separation performance over traditional audio-visual approaches, they do not address speaker generalization, which is a crucial issue in supervised speech separation. In other words, they have been only evaluated in a speaker-dependent way, in which the speakers are not allowed to change from training to testing. Recent studies~\cite{ephrat2018looking},~\cite{afouras2018conversation},~\cite{owens2018audio},~\cite{morrone2019face},~\cite{wu2019time} have developed algorithms for speaker-independent speech separation. Ephrat {\it et al.}~\cite{ephrat2018looking} designed a multi-stream neural network based on dilated convolutions and bidirectional long short-term memory (BLSTM), which leads to significantly better performance than several earlier speaker-dependent models. Afouras {\it et al.}~\cite{afouras2018conversation} utilize two subnetworks to predict the magnitude spectrogram and the phase spectrogram of clean speech, respectively. In~\cite{owens2018audio}, a DNN is trained to predict whether audio and visual streams are temporally synchronized, which is then used to produce multisensory features for speech separation. Wu {\it et al.}~\cite{wu2019time} developed a time-domain audio-visual model for target speaker separation. Note that these studies address monaural speech separation in close-talk scenarios.

In a real-world acoustic environment, speech signals are usually distorted by reverberation from surface reflections. Dereverberation has been actively studied for decades~\cite{avendano1996study},~\cite{naylor2010speech},~\cite{nakatani2010speech},~\cite{hazrati2013blind}. Although deep learning based approaches have significantly improved dereverberation performance in recent years~\cite{han2014learning},~\cite{wu2016reverberation},~\cite{xiao2016speech},~\cite{zhao2019two}, reverberation remains a well-recognized challenge, especially when it is combined with background noise, interfering speech, or both. Despite the promising progress on audio-visual speech separation, few of recent studies deal with both speech separation and dereverberation in a multimodal way. Given the importance of separation and dereverberation to both human and machine listeners (e.g. automatic speech recognition) in noisy and reverberant environments, we address speaker-independent multi-channel speech separation and dereverberation in this study, which aims to separate target speech from interfering speech, background noise and room reverberation. Inspired by recent works~\cite{khan2013two},~\cite{tan2018two},~\cite{zhao2019two} on speech separation, we believe that it is likely more effective to address separation and dereverberation in separate stages due to their intrinsical differences. Hence, we first separate target reverberant speech from interfering speech and background noise using a dilated CNN, and then employ a BLSTM to dereverberate the separated speech signal. Subsequently, the two-stage model is jointly trained to optimize a new multi-objective loss function, which combines a mean squared error (MSE) loss in the T-F domain and a scale-invariant signal-to-noise ratio (SI-SNR) loss in the time domain. Our experimental results show that the proposed multimodal network improves extended short-time objective intelligibility (ESTOI)~\cite{jensen2016algorithm} by 21.10\% and perceptual evaluation of speech quality (PESQ)~\cite{rix2001perceptual} by 0.79 over the unprocessed mixtures. Moreover, we find that the proposed network considerably outperforms several one-stage and two-stage baselines. In this study, audio-visual based joint speech separation and dereverberation are thoroughly investigated in far-field scenarios, where interfering speech, background noise and room reverberation are present.

The rest of this paper is organized as follows. In Section~\ref{sec:signal_model}, we introduce the multi-channel far-field signal model. Section~\ref{sec:features} provides a brief description of several auditory and visual features used in this study. In Section~\ref{sec:alg}, we describe our proposed audio-visual multimodal network architecture in detail. The experimental setup is provided in Section~\ref{sec:exp_setup}. In Section~\ref{sec:exp_results}, we present and discuss experimental results. Section~\ref{sec:conclusion} concludes this paper.

\begin{figure}[t]
	\centering	
	\subfigure[]{%
		\includegraphics[width=8cm]{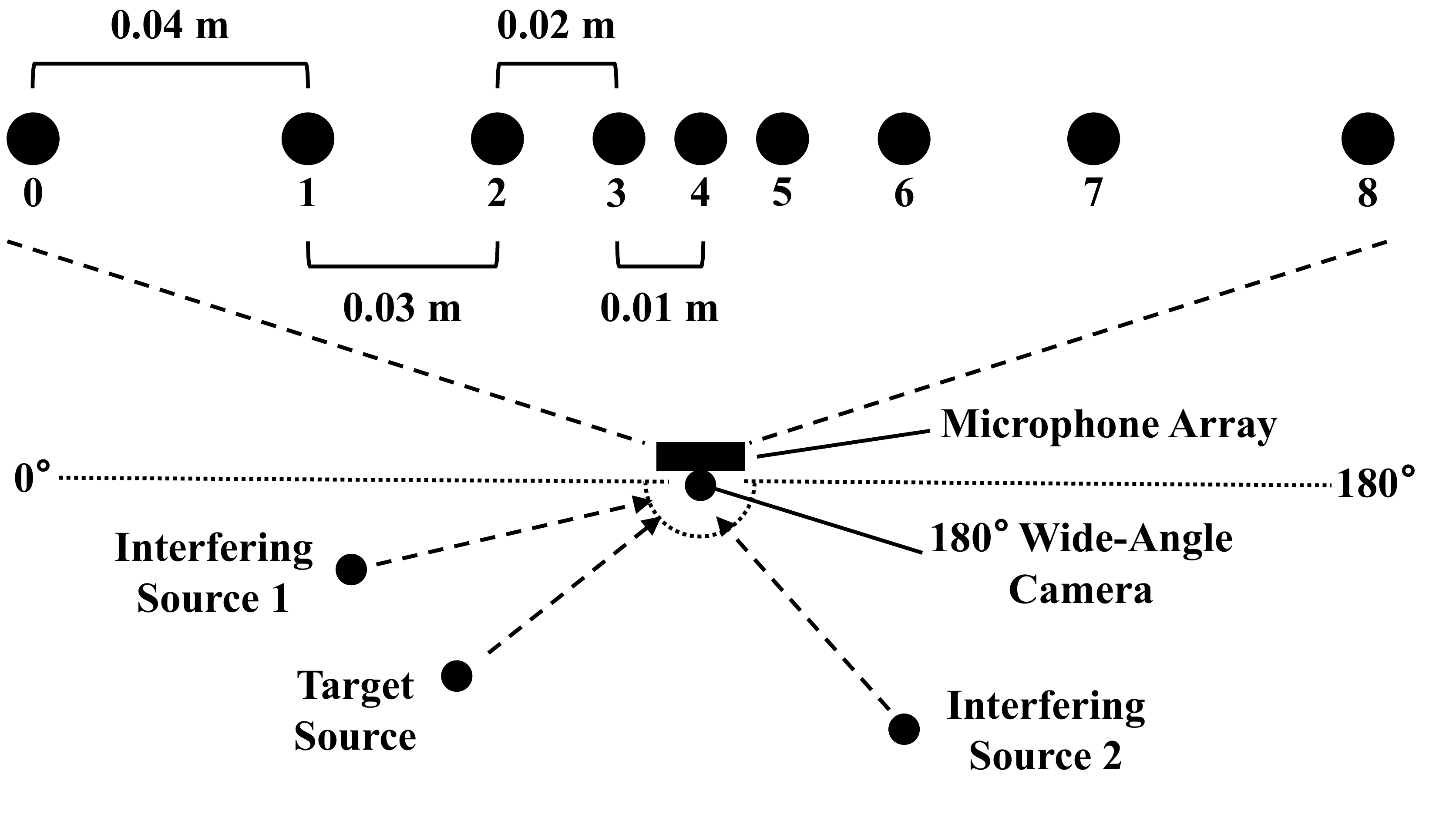}%
		\label{fig:subfig:array_config}%
	}
	
	\subfigure[]{%
		\includegraphics[width=8cm]{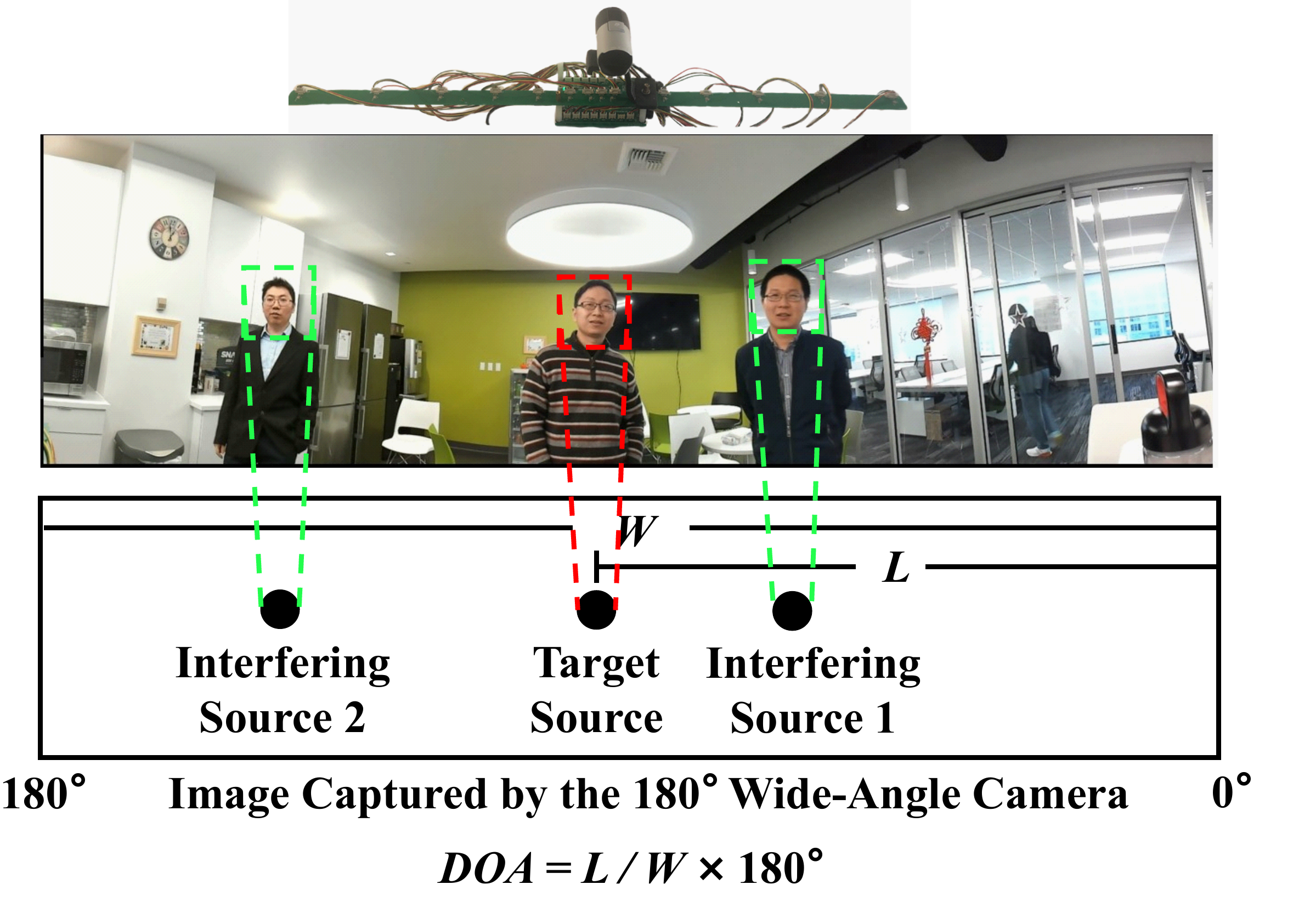}%
		\label{fig:subfig:doa}%
	}%
	
	\caption{(a) Configuration of the linear microphone array, in which nine microphones are nonuniformly spaced and symmetrically distributed. A 180-degree wide-angle camera is co-located with the microphone array center, and the camera is well aligned with the linear array. (b) An example of DOA calculation according to the physical location of the target face in the whole captured video view.}
	\label{fig:array}
\end{figure}

\section{Multi-Channel Far-Field Signal Model}
\label{sec:signal_model}
Let $k$ and $m$ be the time sample index and the channel index, respectively. Thus the far-field speech mixture $y^{(m)}$ can be modeled as 
\begin{equation}
\begin{split}
y^{(m)}[k] &= s[k] * h_s^{(m)}[k] + \sum_{i}{s_i[k] * h_i^{(m)}[k]} \\
&\quad + n[k] * h_n^{(m)}[k],
\end{split}
\end{equation}
where $s$, $s_i$ and $n$ denote the target speech source, the $i$-th interfering speech source and the background noise source, respectively, and $h_s$, $h_i$ and $h_n$ the room impulse responses (RIRs) corresponding to the target speech source, the $i$-th interfering speech source and the background noise source, respectively. The convolution operation is represented by $*$. The objective of this study is to estimate the anechoic target speech signal from the $M$-channel far-field speech mixture $\mathbf{y}=\left[y^{(1)}, y^{(2)}, \dots, y^{(M)}\right]$, as well as the visual streams of the target and interfering speakers' lip images. In this study, we use a linear array of nine microphones as depicted in Fig.~\ref{fig:subfig:array_config}. We number the nine microphones from left to right as 0, 1, $\dots$, 8, respectively. Without loss of generality, we treat the clean speech signal picked up by microphone 0 as the target signal.

\section{Auditory and Visual Features}
\label{sec:features}
In this study, we assume that all signals are sampled at 16~kHz. A 32-ms square-root Hann window is employed to segment a speech signal into a set of time frames, with a 50\% overlap between adjacent frames. We resample the visual streams of face images from all videos to 25 frames-per-second (FPS), where face detection is performed using the tools in the dlib library\footnote{http://dlib.net}. From these preprocessed data, we extract three auditory features and a visual feature for multimodal speech separation and dereverberation.

\begin{figure}[t]
	\centering
	
	\subfigure[LPS]{%
		\includegraphics[width=6cm]{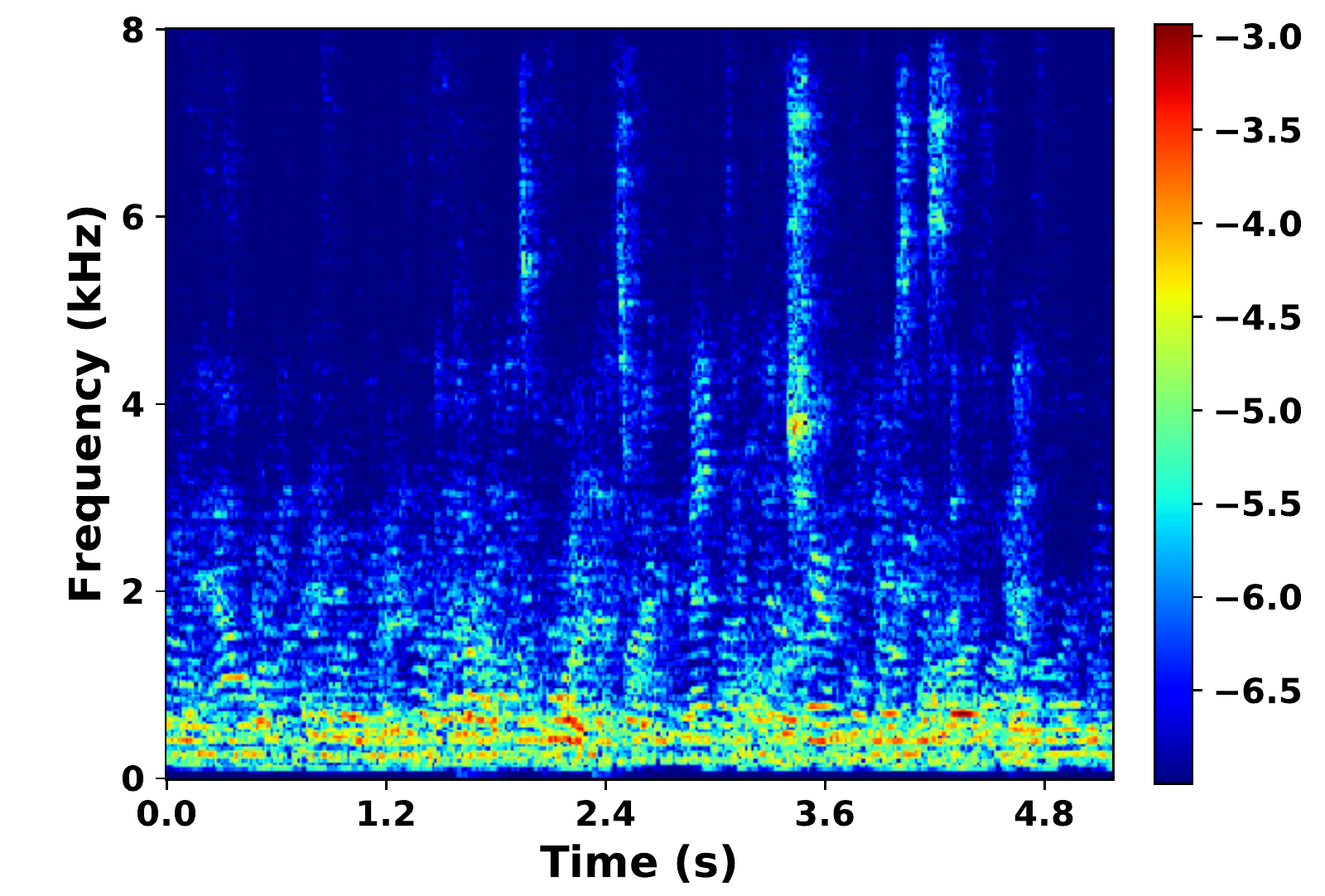}%
		\label{fig:subfig:LPS}%
	}%

	\subfigure[IPD (wrapped into {$\left[-\pi,\pi\right]$})]{%
	\includegraphics[width=6cm]{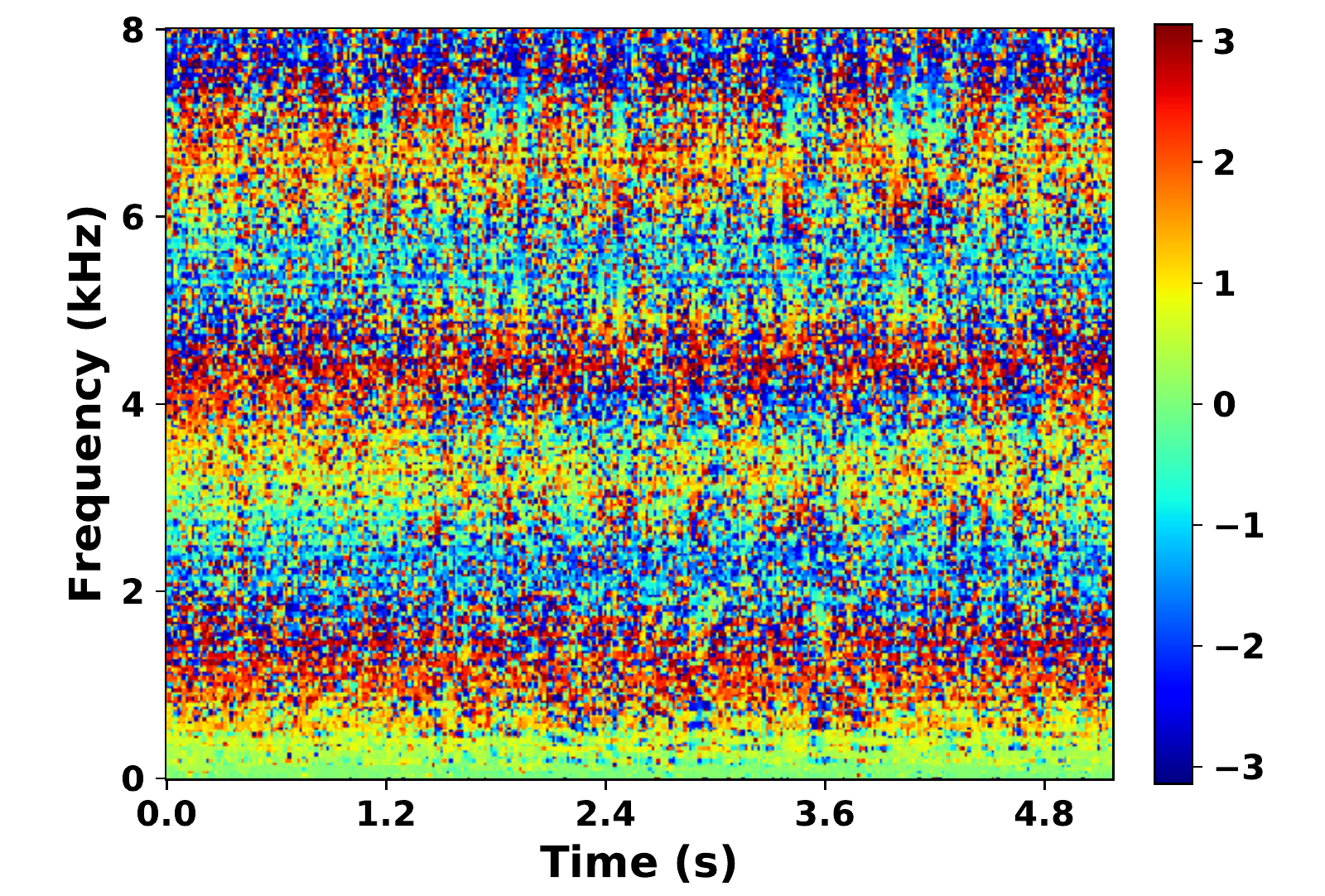}%
	\label{fig:subfig:IPD}%
	}

	\subfigure[cosIPD]{%
	\includegraphics[width=6cm]{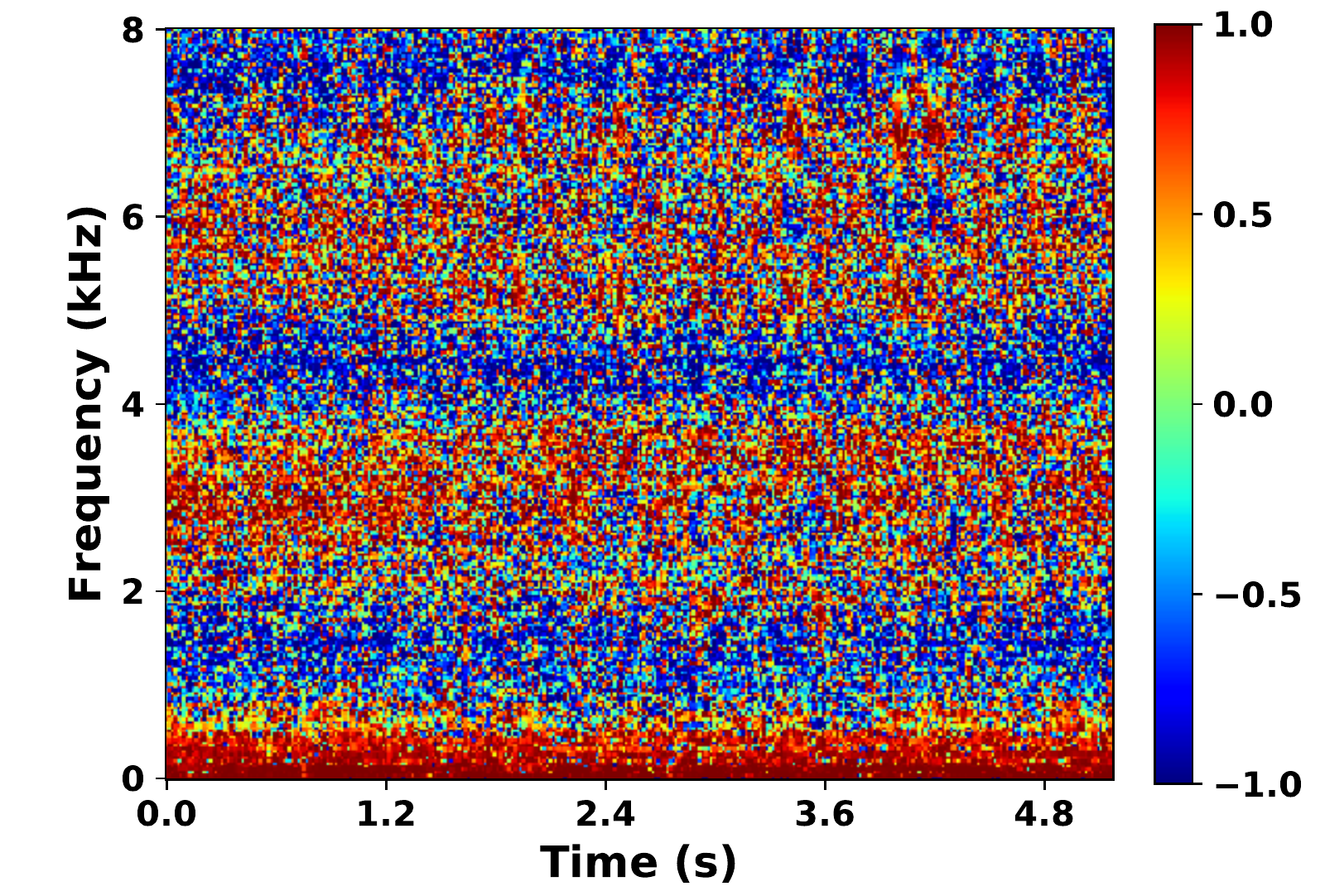}%
	\label{fig:subfig:cosIPD}%
	}%
	
	\caption{(Color Online). Illustration of LPS, IPD and cosIPD features.}
	\label{fig:features}
\end{figure}

\subsection{Log-Power Spectrum}
The log-power spectrum (LPS) of the noisy mixture received by microphone 0, which is a standard spectral representation, is computed on a 512-point short-time Fourier transform (STFT), leading to 257-dimensional (257-D) LPS features. An example of LPS features is shown in Fig.~\ref{fig:subfig:LPS}.

\subsection{Interchannel Phase Difference}
The interchannel phase difference (IPD) is an informative spatial cue which can reflect subtle changes in the direction-of-arrival (DOA) of a sound source. Given a pair of channels $m_1$ and $m_2$, the IPD is defined as
\begin{equation}
\phi_{t,f} = \angle{Y_{t,f}^{(m_2)}} - \angle{Y_{t,f}^{(m_1)}},
\end{equation}
where $Y_{t,f}^{(m_1)}$ and $Y_{t,f}^{(m_2)}$ are the STFT values of the noisy mixture in the T-F unit at time frame $t$ and frequency bin $f$. In this study, we exploit the cosine value of the interchannel phase difference (cosIPD), i.e.
\begin{equation}
\cos \phi_{t,f} = \cos \left(\angle{Y_{t,f}^{(m_2)}} - \angle{Y_{t,f}^{(m_1)}}\right).
\end{equation}
Specifically, we concatenate the cosIPDs between five pairs of channels, i.e. (0, 8), (0, 4), (1, 4), (4, 6) and (4, 5), corresponding to five different microphone distances (see Fig.~\ref{fig:subfig:array_config} for the microphone numbering). The IPD and cosIPD features are illustrated in Figs.~\ref{fig:subfig:IPD} and~\ref{fig:subfig:cosIPD}, respectively.

\begin{figure*}[t]
	\centering
	\includegraphics[width=18cm]{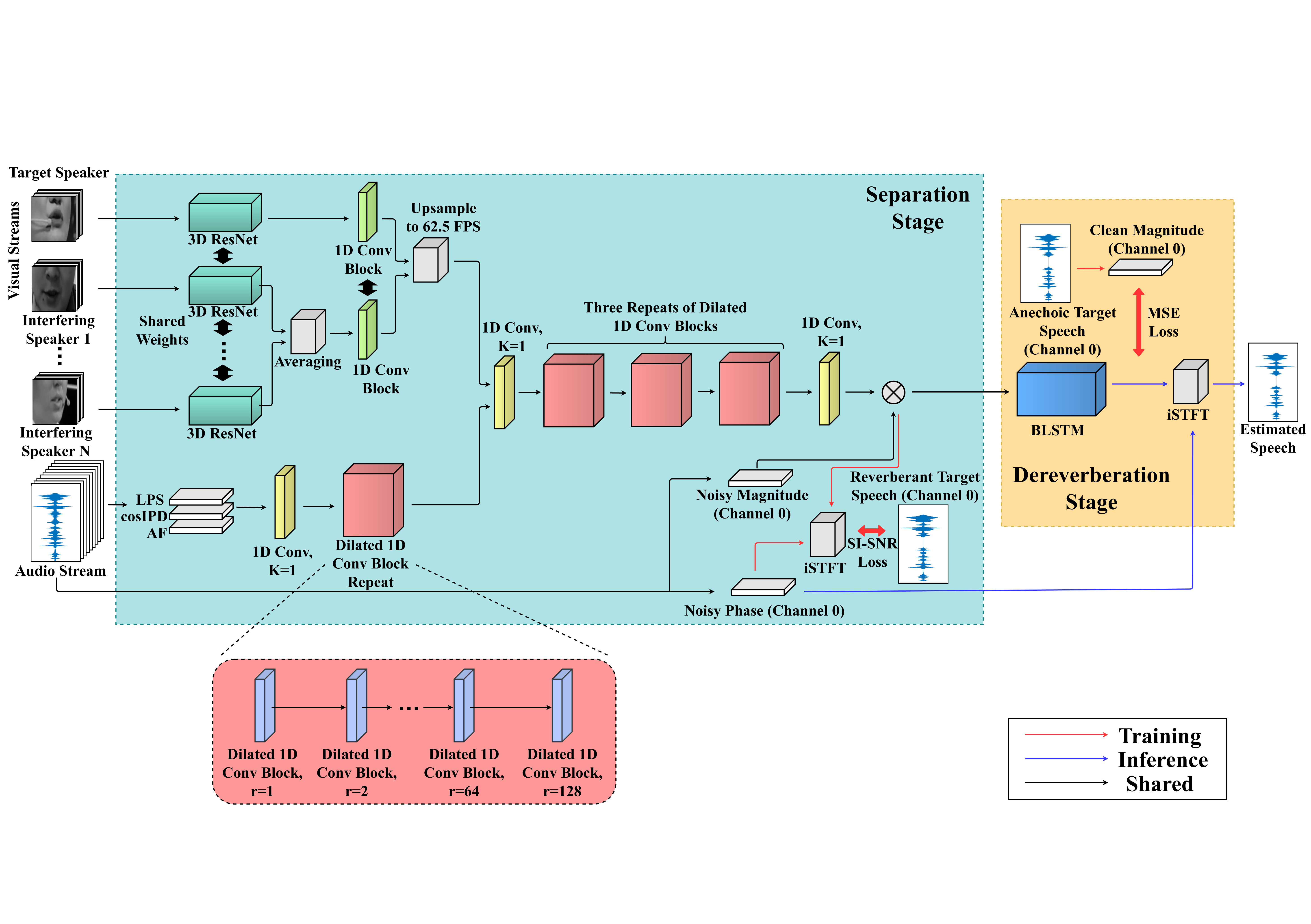}
	\caption{(Color Online). Our proposed multimodal network architecture for joint separation and dereverberation, where $K$ denotes the kernel size and $r$ the dilation rate. The element-wise multiplication is represented by $\bigotimes$. The processes indicated by red lines occur only during training; those indicated by blue lines occur only during inference; those indicated by black lines occur during both.}
	\label{fig:architecture}
\end{figure*}

\subsection{Angle Features}
\label{subsec:af}
We derive an angle feature (AF) for the target speaker, which was first developed in~\cite{chen2018multi}: 
\begin{equation}
A_{t,f} = \sum_{m=0}^{M-1} \frac{\langle e_{f}^{(m)}, {\frac{Y_{t,f}^{(m)}}{Y_{t,f}^{(0)}}} \rangle}{\| e_{f}^{(m)} \| \cdot \| {\frac{Y_{t,f}^{(m)}}{Y_{t,f}^{(0)}}} \|},
\end{equation}
where $M$ represents the number of microphones and $e_{f}^{(m)}$ the steering vector coefficient for the target speaker's DOA at channel $m$ and frequency bin $f$. The inner product is denoted by $\left\langle \cdot, \cdot \right\rangle$, and the vector norm by $\left\| \cdot \right\|$. Note that both $e_{f}^{(m)}$ and $Y_{t,f}^{(m)} / Y_{t,f}^{(0)}$ are complex-valued, and they are treated as 2-D vectors in the operations $\left\langle \cdot, \cdot \right\rangle$ and $\left\| \cdot \right\|$, where their real and imaginary parts are regarded as two vector components. The steering vector is calculated based on the geometry of the microphone array and the arrival direction of the target speech signal, which can be obtained by tracking the target speaker's face from a video captured by a 180-degree wide-angle camera that is co-located with the microphone array center, as shown in Fig.~\ref{fig:subfig:array_config}. The 180-degree wide-angle camera is aligned with the linear microphone array, so that the DOA can be easily calculated as in Fig.~\ref{fig:subfig:doa}. In our experiments, we simulate visual data rather than collect visual data using a real camera.

\subsection{Lip Features}
Each frame of the facial visual stream is cropped to the size of 112$\times$112 based on the mouth region, which amounts to a visual stream of lip images. These images are converted to grayscale using the tools in OpenCV\footnote{https://opencv.org/}. The visual streams from all detected speakers, including both the target speaker and the interfering speakers, are passed into the multimodal network. Note that the faces of the multiple speakers are simultaneously detected by the camera, any of which can be treated as the target speaker, defined by the user. Thus the corresponding target face is used to determine the arrival direction of target speech for the angle feature computation. From an alternative perspective, the lip features and the angle features select the speaker to be separated and allow the network to attend the speech signal coming from the direction of the target speaker.

\section{A Two-Stage Multimodal Network for Joint Separation and Dereverberation}
\label{sec:alg}
In this section, we elaborately describe our proposed two-stage multimodal network architecture for joint separation and dereverberation, which comprises two modules, i.e. a separation module and a dereverberation module. The proposed architecture is illustrated in Fig.~\ref{fig:architecture}.

\begin{figure}[t]
	\centering
	
	\subfigure[]{%
		\includegraphics[width=4.4cm]{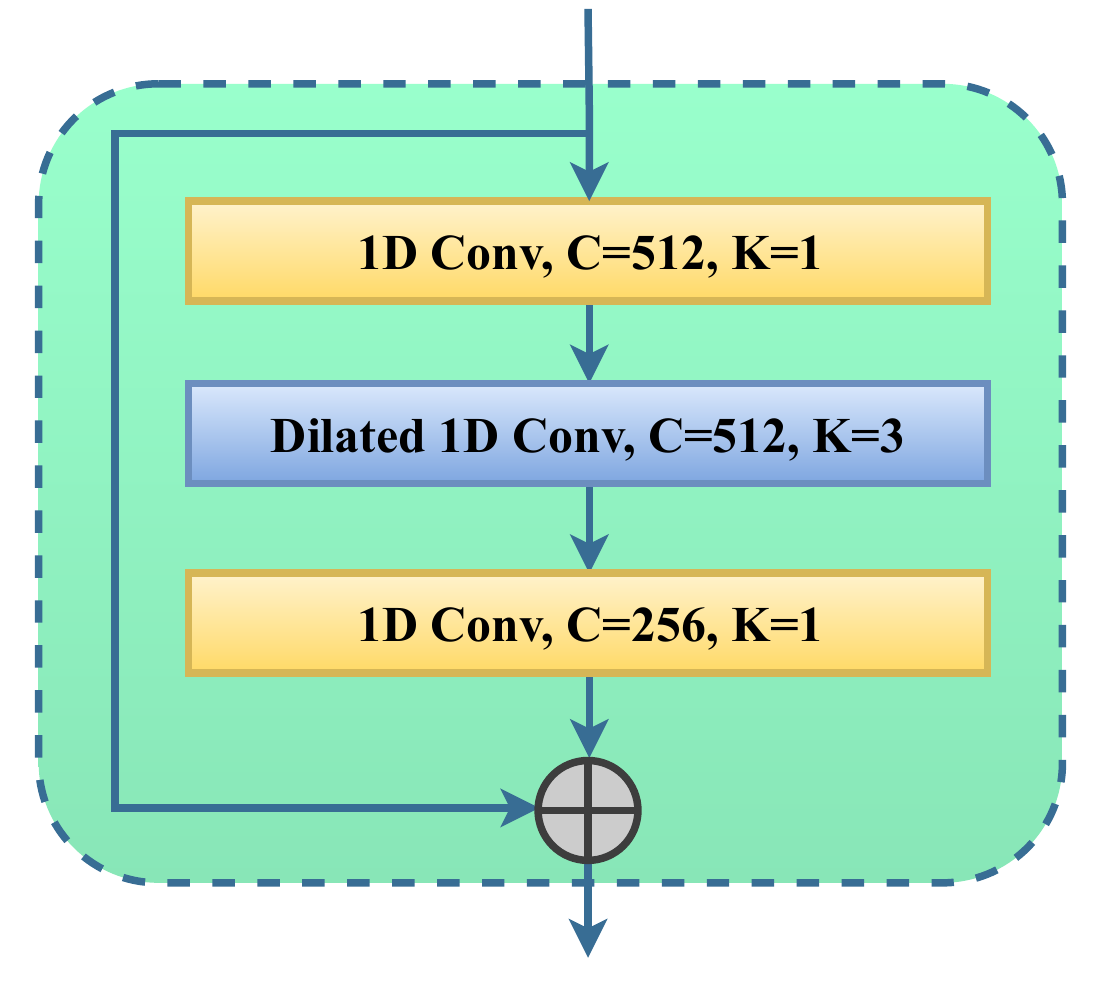}%
		\label{fig:subfig:dilated_block}%
	}%
	\subfigure[]{%
		\includegraphics[width=4.4cm]{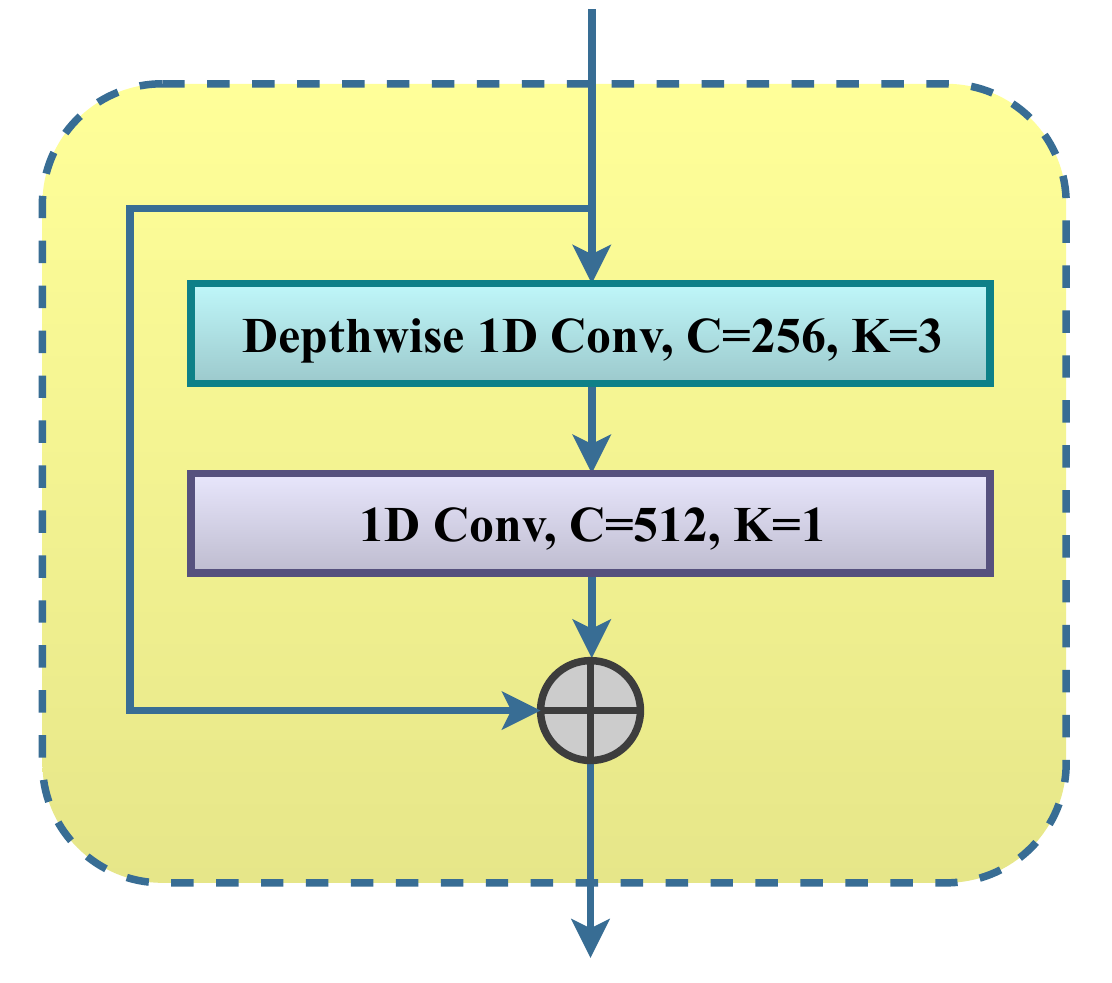}%
		\label{fig:subfig:1d_block}%
	}%
	
	\caption{(Color Online). Diagram of the dilated convolutional block in the audio part (a) and the 1-D convolutional block in the visual part (b), where $\bigoplus$ represents the element-wise summation and $C$ the number of kernels. In (a), the middle layer performs a dilated depthwise convolution. In (b), the first layer performs a depthwise convolution. The first two layers in (a), as well as the first layer in (b), are coupled with batch normalization~\cite{ioffe2015batch} and parametric rectified linear units (PReLUs)~\cite{he2015delving}.}
	\label{fig:conv_blocks}
\end{figure}

\subsection{Separation Stage}
In the separation stage, the LPS features calculated from channel 0 are first passed through a layer normalization~\cite{ba2016layer} layer. The normalized LPS features, the cosIPD features and the angle features are then concatenated into a sequence of 1799-D feature vectors. Subsequently, the sequence is fed into a 1-D convolutional layer with 256 kernels of size 1 (i.e. a pointwise convolutional layer) for dimension reduction. A stack of eight successive 1-D convolutional blocks with dilation rates 1, 2, $\dots$, 2$^{\text{7}}$ is then employed to produce a sequence of audio embeddings. The dilated convolutional block is depicted in Fig.~\ref{fig:subfig:dilated_block}.

The visual stream of a speaker is fed into a spatio-temporal residual network developed in~\cite{afouras2018conversation}, which comprises a 3-D convolutional layer followed by a 18-layer ResNet~\cite{he2016deep}. The spatio-temporal residual networks corresponding to different speakers share weights with one another. As shown in Fig.~\ref{fig:architecture}, the output for the target speaker, as well as the element-wise average of the outputs for all interfering speakers, is passed into a 1-D convolutional block (see Fig.~\ref{fig:subfig:1d_block}) to yield two sequences of visual embeddings. Akin to the spatio-temporal residual networks, the 1-D convolutional block for the target speaker and that for the interfering speakers share weights. The two sequences produced by the 1-D convolutional blocks are concatenated into a sequence of 1024-D embedding vectors, which is subsequently upsampled temporally to 62.5 FPS (=16000$/$(50\%$\times$0.032$\times$16000)) to fit the frame rate of the audio embeddings. The averaging operation across the outputs of the spatio-temporal residual networks for different interfering speakers allows the multimodal network to accept visual streams from an arbitrary number of interfering speakers. In other words, our multimodal network is independent of the number of interfering speakers, unlike the network developed in~\cite{ephrat2018looking}, which can only be used for a fixed number of speakers. Note that when no interfering speaker is detected, we use an all-zero ``visual stream'' as the input of the spatio-temporal residual network for the interfering speaker branch. It should be pointed out that an alternative way is to only use the visual stream of the target speaker as in~\cite{wu2019time}, while additionally using those of interfering speakers potentially leads to more robust performance, particularly when the lip images of the target speaker is blurred or the camera only captures the side face of the target speaker.

We refer to the two network branches that produce the audio and visual embeddings as the audio submodule and the visual submodule, respectively. The audio and visual embeddings are concatenated and then fed into a 1-D pointwise convolutional layer with 256 kernels for audio-visual feature fusion and dimension reduction. Subsequently, the learned high-level features are passed into three repeats of the dilated convolutional blocks (see Fig.~\ref{fig:subfig:dilated_block}) to model temporal dependencies. The dilation rates of eight stacking convolutional blocks within each repeat are assigned with exponentially increasing values, i.e. 1, 2, $\dots$, 2$^{\text{7}}$, which exponentially expand the receptive fields in the time direction, allowing for temporal context aggregation that facilitates estimation. Such a design is originally inspired by the WaveNet for speech synthesis~\cite{vanwavenet} and has been successfully applied to speech separation in recent studies~\cite{rethage2018wavenet},~\cite{tan2018gated},~\cite{tan2019gated},~\cite{luo2019conv}. A 1-D pointwise convolutional layer with rectified linear units (ReLUs) is employed to estimate a ratio mask, which is then element-wise multiplied by the magnitude spectrogram of the noisy mixture from channel 0 to produce that of separated reverberant speech.

During training, the estimated magnitude is combined with noisy phase (from channel 0) to resynthesize a time-domain signal via an inverse short-time Fourier transform (iSTFT). The separation network is trained to maximize the SI-SNR, which has been commonly used as an evaluation metric for speaker separation in recent studies~\cite{isik2016single},~\cite{luo2018speaker},~\cite{luo2019conv},~\cite{wang2019deep}. Thus an SI-SNR loss function can be defined as
\begin{equation}\
\begin{split}
\mathcal{L}_{\text{SI-SNR}} &= -\text{SI-SNR} \\
&= -20\log_{10}{\frac{\|\alpha \cdot \mathbf{s}\|}{\|\hat{\mathbf{s}} - \alpha \cdot \mathbf{s}\|}},
\end{split}
\label{eq:sisnr1}
\end{equation}
where $\mathbf{s} \in \mathbb{R}^{1 \times T}$ and $\hat{\mathbf{s}} \in \mathbb{R}^{1 \times T}$ denote the ground-truth target signal (i.e. reverberant target speech in this stage) and the estimated signal with $T$ time samples, respectively, and $\alpha$ a scaling factor defined as
\begin{equation}
\alpha = \frac{\left\langle \hat{\mathbf{s}}, \mathbf{s} \right\rangle}{\| \mathbf{s} \|^2}.
\end{equation}
Note that $\hat{\mathbf{s}}$ and $\mathbf{s}$ are normalized to zero-mean prior to the calculation to ensure scale invariance. 

\subsection{Dereverberation Stage}
After the attenuation of interfering speech and background noise, the original problem reduces to single-channel speech dereverberation, i.e. recovering anechoic target speech from reverberant target speech estimated by the separation module. In this stage, we employ a BLSTM network with four hidden layers to perform spectral mapping, which takes the spectral magnitudes estimated by the separation module as the input. The reason for using spectral mapping rather than ratio masking in this stage is two-fold. First, ratio masking is well justified for separation under the assumption that target speech and background interference are uncorrelated, which holds well for additive noise (including background noise and interfering speech) but not for convolutive interference as in the case of reverberation~\cite{wang2018supervised}. Second, speech separation algorithms commonly introduce processing artifacts into the target speech signal~\cite{jahn2016wide},~\cite{wang2019bridging}. It is likely difficult for ratio masking to suppress such processing artifacts introduced by the separation module, particularly considering that these artifacts are correlated with the target speech signal.

During training, the well-trained parameters in the separation module are frozen, and those in the dereverberation module are trained to optimize an MSE loss function, which compares the ground-truth magnitude spectrogram $|S|$ with the estimated magnitude spectrogram $|\hat{S}|$:
\begin{equation}
\mathcal{L}_{\text{MSE}} = \mathbb{E}\left[(| \hat{S}_{t,f} | - | S_{t,f} |)^2\right],
\label{eq:MSE}
\end{equation}
where $\mathbb{E}$ represents an averaging operation over all T-F units of all training samples within a minibatch. The complex modulus is denoted by $\left| \cdot \right|$, i.e. the absolute value of a complex number. The use of the MSE loss, rather than the SI-SNR loss, is motivated by the observation that training with the SI-SNR loss leads to far slower convergence and worse performance on speech dereverberation than on speaker separation, which is likely because the SI-SNR loss is based on a sample-wise error in the time domain and thus sensitive to the highly correlated structure between the direct sound and the reverberations~\cite{luo2018real}.

\subsection{Joint Training with a Multi-Objective Loss Function}
\label{subsec:JT}
After the two modules are well trained separately, we treat them as an integrated network for joint training (JT). Akin to the dereverberation stage, a straightforward way to train the network is to optimize an MSE loss that is calculated on the output of the dereverberation module, as shown in Eq.~(\ref{eq:MSE}). Unlike the SI-SNR loss, however, the MSE loss only reflects the difference between the target magnitude and the estimated magnitude, where the phase remains unaddressed. A recent study~\cite{paliwal2011importance} suggests that considerable improvements in both objective and subjective speech quality can be achieved by accurate phase spectrum estimation, which implies the importance of dealing with the phase to producing high-quality separated speech. Since the SI-SNR loss is calculated in the time domain, both the magnitude error and the phase error are incorporated. In other words, training with the SI-SNR loss implicitly involves phase estimation.

\begin{figure*}[t]
	\centering
	\includegraphics[width=16cm]{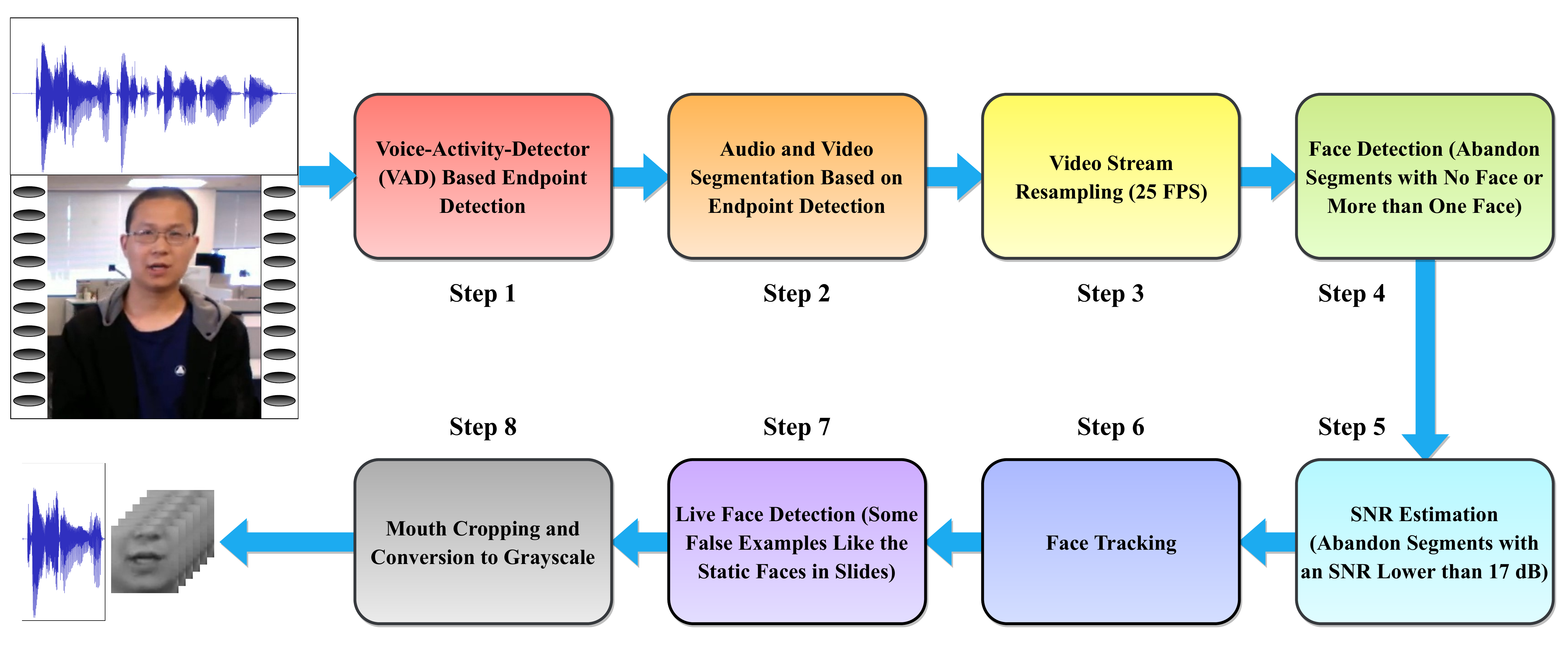}
	\caption{(Color Online). Diagram of our dataset creation pipeline. In Steps 4 and 5, data filtering is performed based on the number of detected faces and the estimated SNR, respectively.}
	\label{fig:data_prepare}
\end{figure*}

Motivated by this fact, we design a multi-objective loss function for joint training, which combines the MSE loss and the SI-SNR loss:
\begin{equation}
\mathcal{L}_{\text{Multi-Obj}} = \mathcal{L}_{\text{MSE}} + \lambda \cdot \mathcal{L}_{\text{SI-SNR}},
\end{equation}
where $\lambda$ is a pre-defined weighting factor. However, such a combination of the two losses is dubious, as the MSE loss is guaranteed to be non-negative while the SI-SNR loss in Eq.~(\ref{eq:sisnr1}) is unbounded. Specifically, there are two critical flaws in this design. First, it is tricky to choose an appropriate value of $\lambda$, which weights a loss $\mathcal{L}_{\text{SI-SNR}}$ with an uncertain sign. Second, when $\mathcal{L}_{\text{MSE}}$ is close to $-\lambda \cdot \mathcal{L}_{\text{SI-SNR}}$, the multimodal network is discouraged to learn due to the gradients that are close to zero.

In order to mitigate these problems, an intuitive way is to define an alternative SI-SNR loss that is ensured to be non-negative like the MSE loss. Note that Eq.~(\ref{eq:sisnr1}) can be rewritten into
\begin{equation}
\mathcal{L}_{\text{SI-SNR}} = 20\log_{10}{\frac{\|\hat{\mathbf{s}} - \alpha \cdot \mathbf{s}\|}{\|\alpha \cdot \mathbf{s}\|}}.
\end{equation}
Thus we define a new SI-SNR loss as
\begin{equation}
\mathcal{L}'_{\text{SI-SNR}} = 20\log_{10}{\left(\frac{\|\hat{\mathbf{s}} - \alpha \cdot \mathbf{s}\|}{\|\alpha \cdot \mathbf{s}\|} + 1\right)}.
\end{equation}
Therefore, we train the multimodal network with the multi-objective loss function:
\begin{equation}
\mathcal{L}'_{\text{Multi-Obj}} = \mathcal{L}_{\text{MSE}} + \lambda \cdot \mathcal{L}'_{\text{SI-SNR}},
\end{equation}
where $\mathcal{L}_{\text{MSE}} \in [0, +\infty)$ and $\mathcal{L}'_{\text{SI-SNR}} \in [0, +\infty)$. During inference, the estimated spectral magnitude is combined with the noisy phase to recover the time-domain waveform.

\section{Experimental Setup}
\label{sec:exp_setup}
%\subsection{Data Preparation}
We create a new Chinese Mandarin audio-visual dataset for this study. Specifically, we collect roughly 10,000 videos of Chinese Mandarin lectures from YouTube, and then pass them through a dataset creation pipeline, which is shown in Fig.~\ref{fig:data_prepare}. A series of processing steps in the pipeline leads to an audio-visual dataset including approximately 170,000 short video clips with a total duration of around 155 hours. Each video clip in the dataset, which has a duration between 500~ms and 13~s, corresponds to an audio signal (i.e. the soundtrack of the video clip) and a visual stream of grayscale lip (mouth) images.

\begin{figure}[t]
	\centering
	
	\subfigure[]{%
		\includegraphics[width=4.4cm]{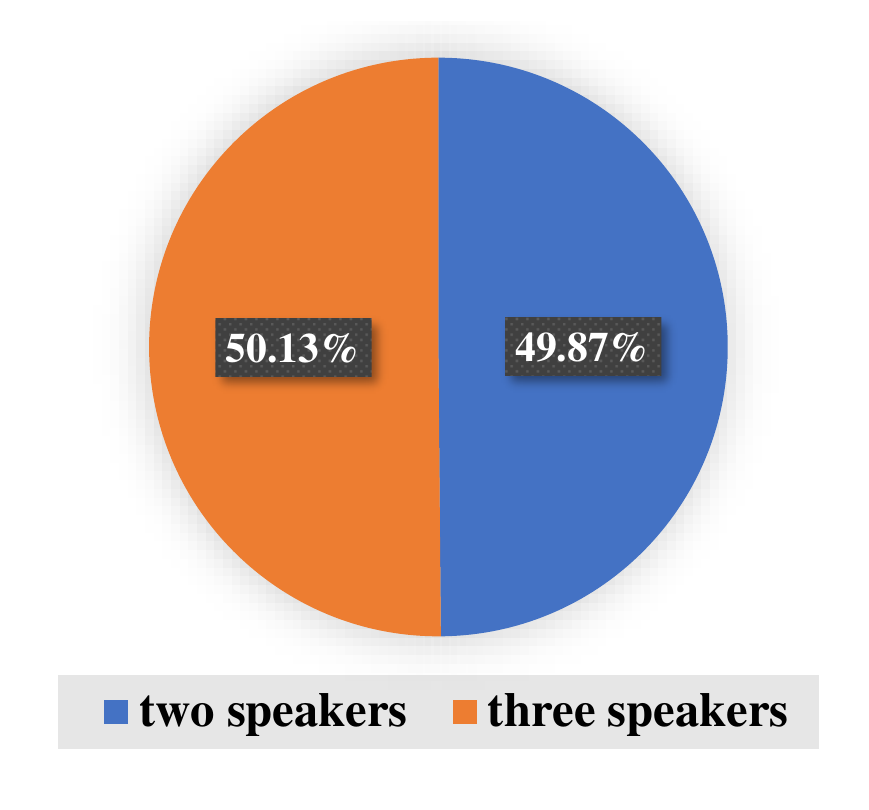}%
		\label{fig:subfig:train_speakers}%
	}%
	\subfigure[]{%
		\includegraphics[width=4.4cm]{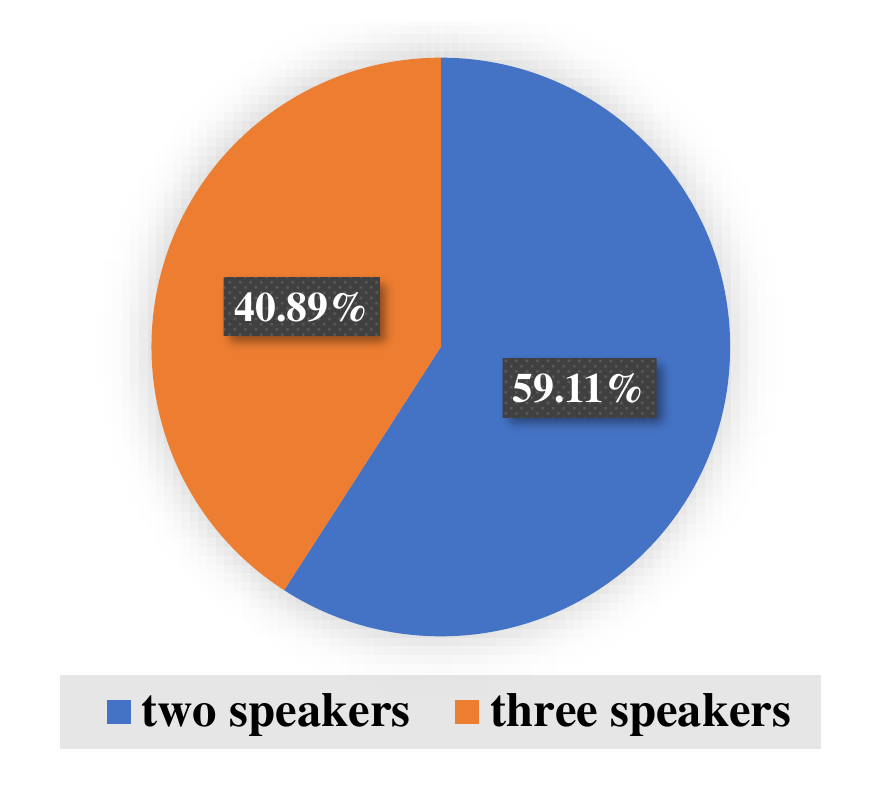}%
		\label{fig:subfig:test_speakers}%
	}%
	
	\subfigure[]{%
		\includegraphics[width=4.4cm]{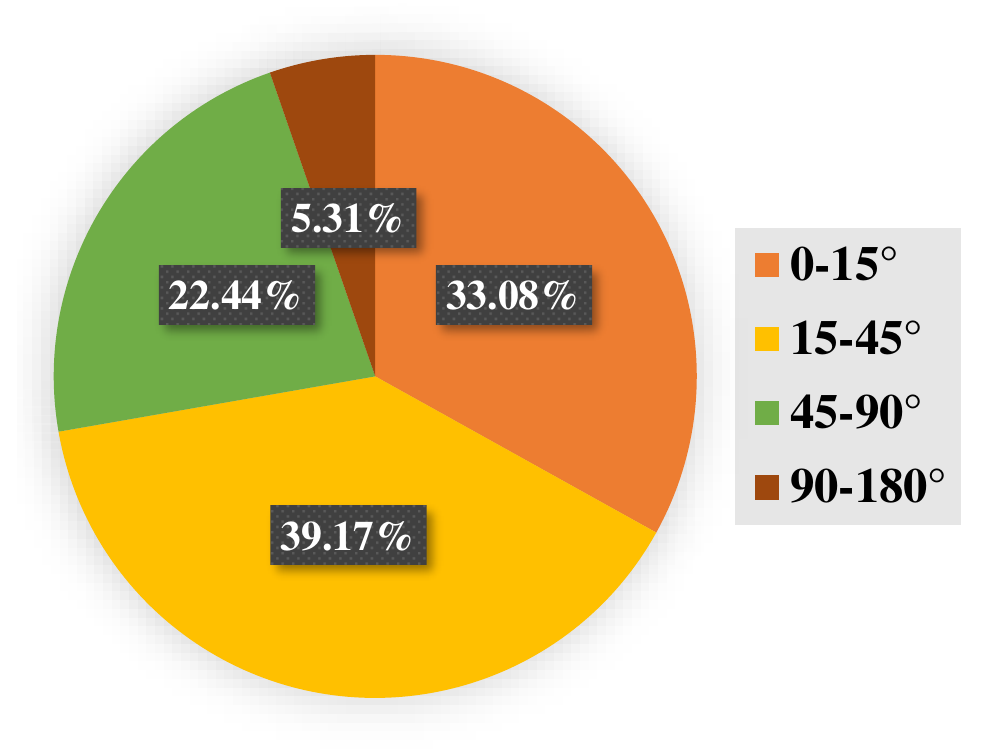}%
		\label{fig:subfig:train_angles}%
	}%
	\subfigure[]{%
		\includegraphics[width=4.4cm]{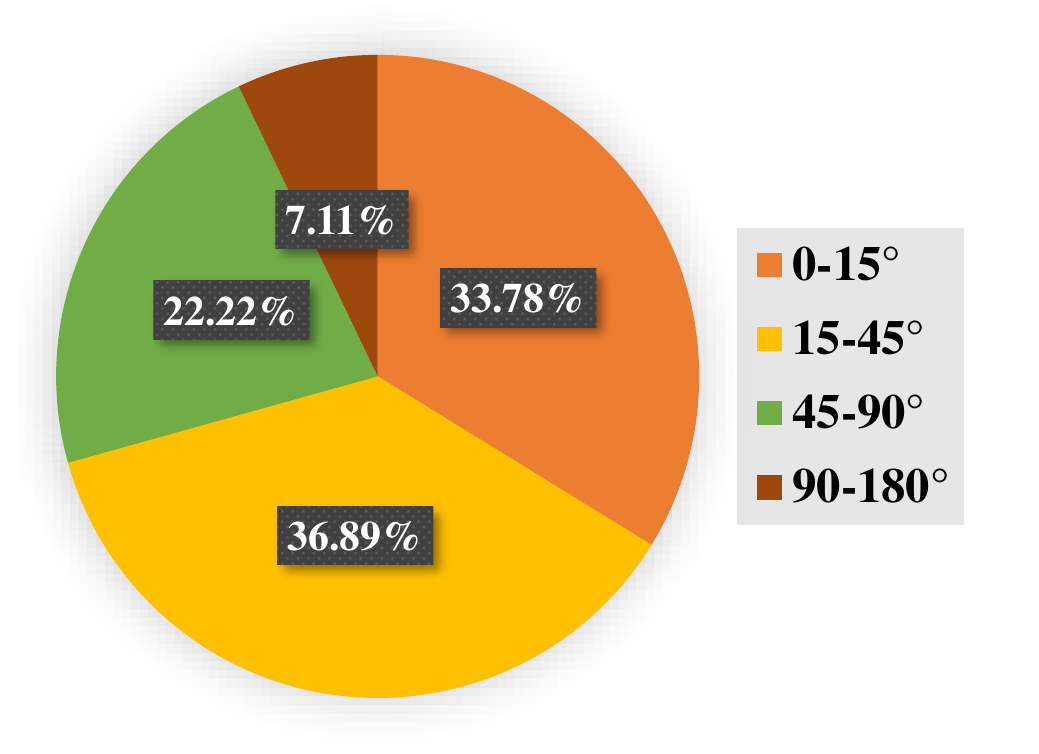}%
		\label{fig:subfig:test_angles}%
	}%
	
	\caption{(Color Online). Distribution of two-speaker and three-speaker mixtures and that of the angle between the DOA's of the target speech signal and an interfering speech signal in the training set (a) (c) and the test set (b) (d).}
	\label{fig:distribution}
\end{figure}

\begin{table*}[t]\scriptsize
	\caption{Comparisons of different approaches in ESTOI and PESQ.} 
	\newcolumntype{V}{!{\vrule width 0.7pt}}
	\newcolumntype{U}{!{\vrule width 0.7pt}}
	\centering
	%\[
	\resizebox{\textwidth}{!}{
		\begin{tabular}{VcVcVcVccccVcVccccVcV}
			\Xhline{0.7pt}
			\textbf{Approaches} & \textbf{IDs} & \textbf{Loss Functions} & \multicolumn{5}{cV}{\textbf{ESTOI (in \%)}} & \multicolumn{5}{cV}{\textbf{PESQ}} \\
			\Xhline{0.7pt}
			\multirow{2}{*}{Angle Ranges} & \multirowcell{2}{-} & \multirowcell{2}{-} & 0-15$^\circ$ & 15-45$^\circ$ & 45-90$^\circ$ & 90-180$^\circ$ & \multirowcell{2}{Avg.} & 0-15$^\circ$ & 15-45$^\circ$ & 45-90$^\circ$ & 90-180$^\circ$ & \multirowcell{2}{Avg.} \\ & & & (34\%) & (37\%) & (22\%) & (7\%) & & (34\%) & (37\%) & (22\%) & (7\%) & \\
			\Xhline{0.7pt}
			Unprocessed & - & - & 42.07 & 44.77 & 46.31 & 45.09 & 44.21 & 1.55 & 1.60 & 1.63 & 1.74 & 1.60 \\
			\Xhline{0.7pt}
			BLSTM & 1 & $\mathcal{L}_{\text{MSE}}$ & 51.09 & 57.84 & 60.88 & 63.76 & 56.63 & 1.79 & 2.00 & 2.07 & 2.18 & 1.96 \\
			Dilated CNN (audio only) & 2 & $\mathcal{L}_{\text{MSE}}$ & 50.10 & 56.62 & 60.41 & 62.14 & 55.63 & 1.73 & 1.91 & 2.01 & 2.11 & 1.88 \\
			Dilated CNN & 3 & $\mathcal{L}_{\text{SI-SNR}}$ & 47.75 & 56.85 & 60.99 & 63.73 & 55.15 & 1.59 & 1.85 & 1.96 & 2.07 & 1.80 \\
			\rowcolor{mygray}
			Dilated CNN & 4 & $\mathcal{L}_{\text{MSE}}$ & 51.81 & 59.32 & 62.59 & 65.88 & 57.95 & 1.84 & 2.05 & 2.15 & 2.25 & 2.01 \\
			\Xhline{0.7pt}
			Dilated CNN+BLSTM (w/o JT) & 5 & $\mathcal{L}_{\text{SI-SNR}}$, $\mathcal{L}_{\text{SI-SNR}}$ & 55.84 & 60.92 & 63.22 & 66.47 & 60.09 & 1.95 & 2.08 & 2.20 & 2.30 & 2.08 \\
			Dilated CNN+BLSTM (w/o JT) & 6 & $\mathcal{L}_{\text{SI-SNR}}$, $\mathcal{L}_{\text{MSE}}$ & 56.40 & 60.83 & 63.27 & 66.75 & 60.28 & 2.10 & 2.20 & 2.30 & 2.40 & 2.20 \\
			Dilated CNN+WPE & 7 & $\mathcal{L}_{\text{SI-SNR}}$~(Separation Stage) & 51.84 & 58.53 & 61.50 & 64.10 & 57.30 & 1.87 & 2.06 & 2.12 & 2.24 & 2.02 \\
			Dilated CNN+BLSTM (One-Stage) & 8 & $\mathcal{L}_{\text{MSE}}$ & 50.44 & 56.32 & 58.91 & 61.56 & 55.26 & 1.81 & 2.02 & 2.10 & 2.21 & 1.98 \\
			Dilated CNN+BLSTM (w/ JT) & 9 & $\mathcal{L}_{\text{SI-SNR}}$, $\mathcal{L}_{\text{MSE}}$, $\mathcal{L}'_{\text{Multi-Obj}}$~($\lambda$=0) & 60.08 & 65.81 & 67.65 & 70.82 & 64.42 & 2.19 & 2.37 & 2.45 & 2.56 & 2.34 \\
			Dilated CNN+BLSTM (w/ JT) & 10 & $\mathcal{L}_{\text{SI-SNR}}$, $\mathcal{L}_{\text{MSE}}$, $\mathcal{L}'_{\text{Multi-Obj}}$~($\lambda$=0.01) & 59.86 & 65.26 & 67.07 & 70.42 & 64.19 & 2.20 & 2.35 & 2.44 & 2.53 & 2.33 \\
			Dilated CNN+BLSTM (w/ JT) & 11 & $\mathcal{L}_{\text{SI-SNR}}$, $\mathcal{L}_{\text{MSE}}$, $\mathcal{L}'_{\text{Multi-Obj}}$~($\lambda$=0.02) & 60.26 & 65.32 & 67.28 & 71.27 & 64.45 & 2.22 & 2.37 & 2.46 & 2.55 & 2.35 \\
			Dilated CNN+BLSTM (w/ JT) & 12 & $\mathcal{L}_{\text{SI-SNR}}$, $\mathcal{L}_{\text{MSE}}$, $\mathcal{L}'_{\text{Multi-Obj}}$~($\lambda$=0.05) & 60.76 & 65.92 & 68.11 & 71.30 & 65.03 & 2.24 & 2.41 & 2.50 & 2.59 & 2.38 \\
			\rowcolor{mygray}
			Dilated CNN+BLSTM (w/ JT) & 13 & $\mathcal{L}_{\text{SI-SNR}}$, $\mathcal{L}_{\text{MSE}}$, $\mathcal{L}'_{\text{Multi-Obj}}$~($\lambda$=0.08) & \textbf{60.73} & \textbf{66.33} & \textbf{68.45} & \textbf{72.20} & \textbf{65.31} & \textbf{2.24} & \textbf{2.42} & \textbf{2.52} & \textbf{2.62} & \textbf{2.39} \\
			Dilated CNN+BLSTM (w/ JT) & 14 & $\mathcal{L}_{\text{SI-SNR}}$, $\mathcal{L}_{\text{MSE}}$, $\mathcal{L}'_{\text{Multi-Obj}}$~($\lambda$=0.1) & 60.78 & 65.81 & 68.15 & 71.74 & 65.03 & 2.22 & 2.39 & 2.49 & 2.59 & 2.37 \\
			Dilated CNN+BLSTM (w/ JT) & 15 & $\mathcal{L}_{\text{SI-SNR}}$, $\mathcal{L}_{\text{MSE}}$, $\mathcal{L}'_{\text{Multi-Obj}}$~($\lambda$=0.2) & 60.49 & 65.54 & 67.95 & 71.29 & 64.76 & 2.19 & 2.36 & 2.48 & 2.56 & 2.34 \\
			Dilated CNN+BLSTM (w/ JT) & 16 & $\mathcal{L}_{\text{SI-SNR}}$, $\mathcal{L}_{\text{MSE}}$, $\mathcal{L}'_{\text{Multi-Obj}}$~($\lambda$=0.4) & 59.64 & 65.51 & 67.59 & 71.58 & 64.40 & 2.20 & 2.37 & 2.49 & 2.57 & 2.35 \\
			\Xhline{0.7pt}
		\end{tabular}
	}
	%\]
	\label{tab:estoi_pesq}
\end{table*}

Based on this new audio-visual dataset, we simulate multi-channel data for multimodal speech separation and dereverberation. The audio signals from different speakers in the Chinese Mandarin dataset are treated as speech sources (either a target source or an interfering source). Moreover, a random cut from 255 noises recorded indoors is treated as a noise source. These sound sources and a microphone array (see Fig.~\ref{fig:subfig:array_config}) are randomly placed in a simulated room, where the distance between a sound source and the microphone array center is limited to the range of 0.5~m to 6~m. To include a wide variety of reverberant environments, we generate a large set of 6,000 room impulse responses (RIRs) using the image method~\cite{allen1979image} in 2,000 different simulated rooms. The room size is randomly sampled in the range of 4~m~$\times$~4~m~$\times$~3~m to 10~m~$\times$~10~m~$\times$~6~m, and the reverberation time ($T_{60}$) in the range of 0.05~s to 0.7~s. The SNR is randomly chosen from 6, 12, 18, 24 and 30~dB, and the target-to-interferer ratio (TIR) from -6, 0 and 6~dB. Here both the SNR and the TIR are defined on reverberant signals:
\begin{align}
\text{SNR} &= 10 \log_{10}{\frac{\sum_{k}{s_{\text{tar}}^2[k]}}{\sum_{k}{n^2[k]}}}~\text{dB}, \\
\text{TIR} &= 10 \log_{10}{\frac{\sum_{k}{s_{\text{tar}}^2[k]}}{\sum_{k}{s_{\text{int}}^2[k]}}}~\text{dB},
\end{align}
where $s_{\text{tar}}$, $s_{\text{int}}$ and $n$ denote reverberant target speech, reverberant interfering speech and reverberant noise, respectively. Based on the signal model described in Section~\ref{sec:signal_model}, we create roughly 45,000, 200 and 500 mixtures in the training set, the validation set and the test set, respectively. Note that all test speakers and noises are excluded from the training set and the validation set. In other words, we evaluate the models in a speaker- and noise-independent way.

Both the training set and the test set include two-speaker mixtures and three-speaker mixtures, of which the distributions are shown in Figs.~\ref{fig:subfig:train_speakers} and~\ref{fig:subfig:test_speakers}, respectively. Moreover, the distributions of the angle between the DOA's of the target speech signal and an interfering speech signal are shown in Figs.~\ref{fig:subfig:train_angles} and~\ref{fig:subfig:test_angles}. In the case of three speakers (i.e. two interfering speakers), we choose the smaller angle from the two alternatives for counting in Figs.~\ref{fig:subfig:train_angles} and~\ref{fig:subfig:test_angles}, i.e. $\text{Angle}_{\text{DOA}} = \min{\{ \angle{(\text{DOA}_{\text{tar}}, \text{DOA}_{\text{int1}})}, \angle{(\text{DOA}_{\text{tar}}, \text{DOA}_{\text{int2}})} \}}$.

%\subsection{Training Methodology}
All models are trained on 4-second audio-visual chunks using the Adam optimizer~\cite{kingma2015adam} with a learning rate of 0.0002. The minibatch size is set to 20 at the chunk level. The best models are selected by cross validation. In the two-stage approaches, the spectral magnitudes produced by the separation module are normalized via a layer normalization operation prior to being fed into the BLSTM layers. On top of the four stacking BLSTM layers, a fully connected layer with ReLU nonlinearity is used to estimate the spectral magnitudes of anechoic target speech. Specifically, from the input layer to the output layer, the BLSTM has 257, 512, 512, 512, 512, and 257 units, respectively. For joint training, we empirically choose 0.01, 0.02, 0.05, 0.08, 0.1, 0.2 and 0.4 as the value of $\lambda$ for weighing the importance of the MSE loss and the SI-SNR loss.

In this study, we mainly use two metrics to evaluate the models, i.e. ESTOI and PESQ. ESTOI is an improved version of short-time objective intelligibility (STOI)~\cite{taal2011algorithm}, which is an objective speech intelligibility estimator that is commonly used to evaluate the performance of speech enhancement. Specifically, STOI does not highly correlate with subjective listening test results if the target speech signal is distorted by an additive noise source with strong temporal modulations, e.g. a competing speaker~\cite{jensen2016algorithm}. In contrast, ESTOI performs well in such situations, as well as the situations where STOI performs well. Moreover, PESQ is a speech quality estimator that is designed to predict the mean opinion score of a speech quality listening test for certain degradations. The STOI score is typically between 0 and 1, and the PESQ score between -0.5 and 4.5. For both metrics, higher scores indicate better performance.

\section{Experimental Results and Analysis}
\label{sec:exp_results}
\subsection{Results and Comparisons}
Table~\ref{tab:estoi_pesq} presents comprehensive comparisons of different approaches in ESTOI and PESQ. The numbers represent the averages over the test samples in each test condition. We first compare four one-stage baselines with IDs 1-4 (see Table~\ref{tab:estoi_pesq} for the IDs), where separation and dereverberation are jointly performed in a single stage. These approaches treat the anechoic target speech signal as the desired signal during training. In approaches 3 and 4, a dilated CNN with an architecture of the separation module in Fig.~\ref{fig:architecture} is employed to separate target speech from interfering speech, background noise and room reverberation. Specifically, approach 3 trains the dilated CNN with $\mathcal{L}_{\text{SI-SNR}}$, and approach 4 with $\mathcal{L}_{\text{MSE}}$. As shown in Table~\ref{tab:estoi_pesq}, approach 4 consistently outperforms approach 3 in both ESTOI and PESQ, which suggests that $\mathcal{L}_{\text{MSE}}$ is more advantageous than $\mathcal{L}_{\text{SI-SNR}}$ in the presence of room reverberation. In approach 2, we remove the visual submodule from the baseline in approach 4, which leads to an audio-only baseline. One can observe that approach 4 yields consistently higher ESTOI and PESQ than approach 2, which demonstrates the usefulness of visual inputs and thus the effectiveness of multimodal separation and dereverberation. Approach 1 (simply denoted as ``BLSTM'') uses a BLSTM model, which takes the raw LPS, cosIPD and AF features, as well as the visual embeddings produced by the visual submodule, as inputs. It has four BLSTM hidden layers with 512 units in each layer, and a fully connected layer with ReLUs is used to estimate a ratio mask. From Table~\ref{tab:estoi_pesq}, we can see that approach 4 produces higher ESTOI and PESQ than approach 1.

\begin{figure}[t]
	\centering
	\includegraphics[width=9cm]{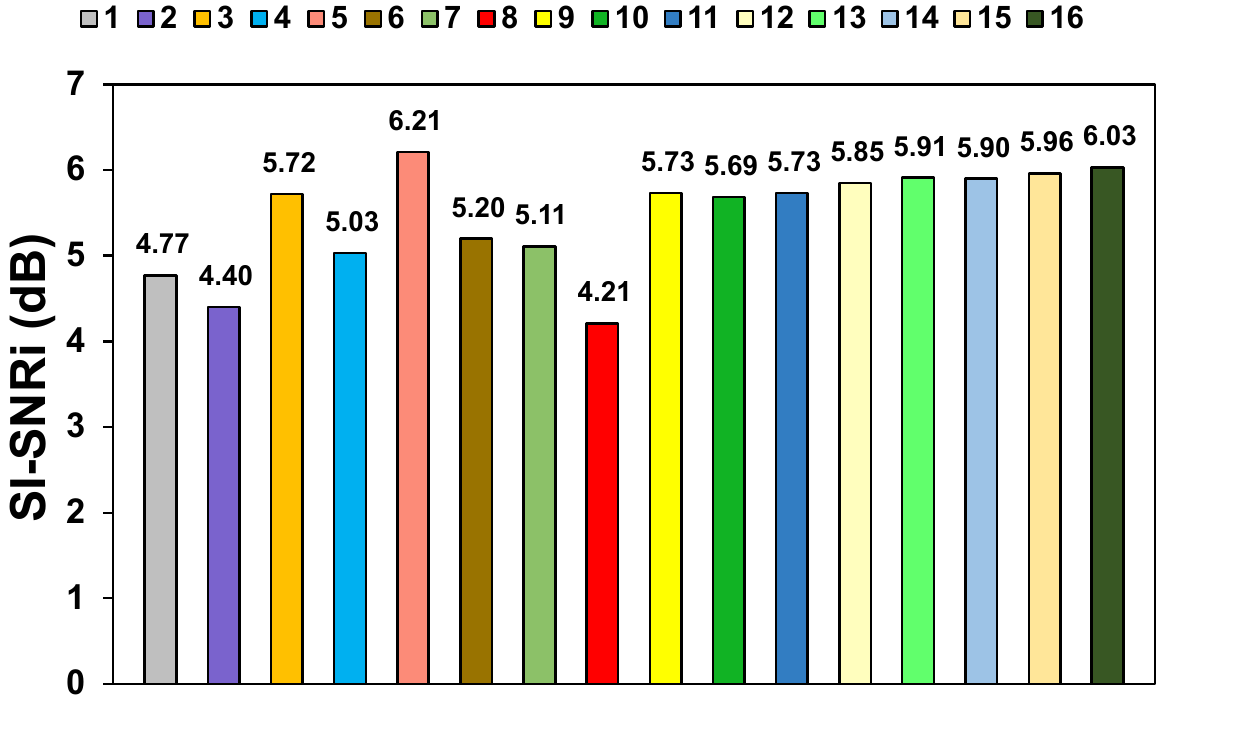}
	\caption{(Color Online). SI-SNRi in dB over the unprocessed mixtures. See Table~\ref{tab:estoi_pesq} for the IDs of different approaches.}
	\label{fig:si-snri}
\end{figure}

\begin{figure}[t]
	\centering
	
	\subfigure[Unprocessed Mixture]{%
		\includegraphics[width=8cm]{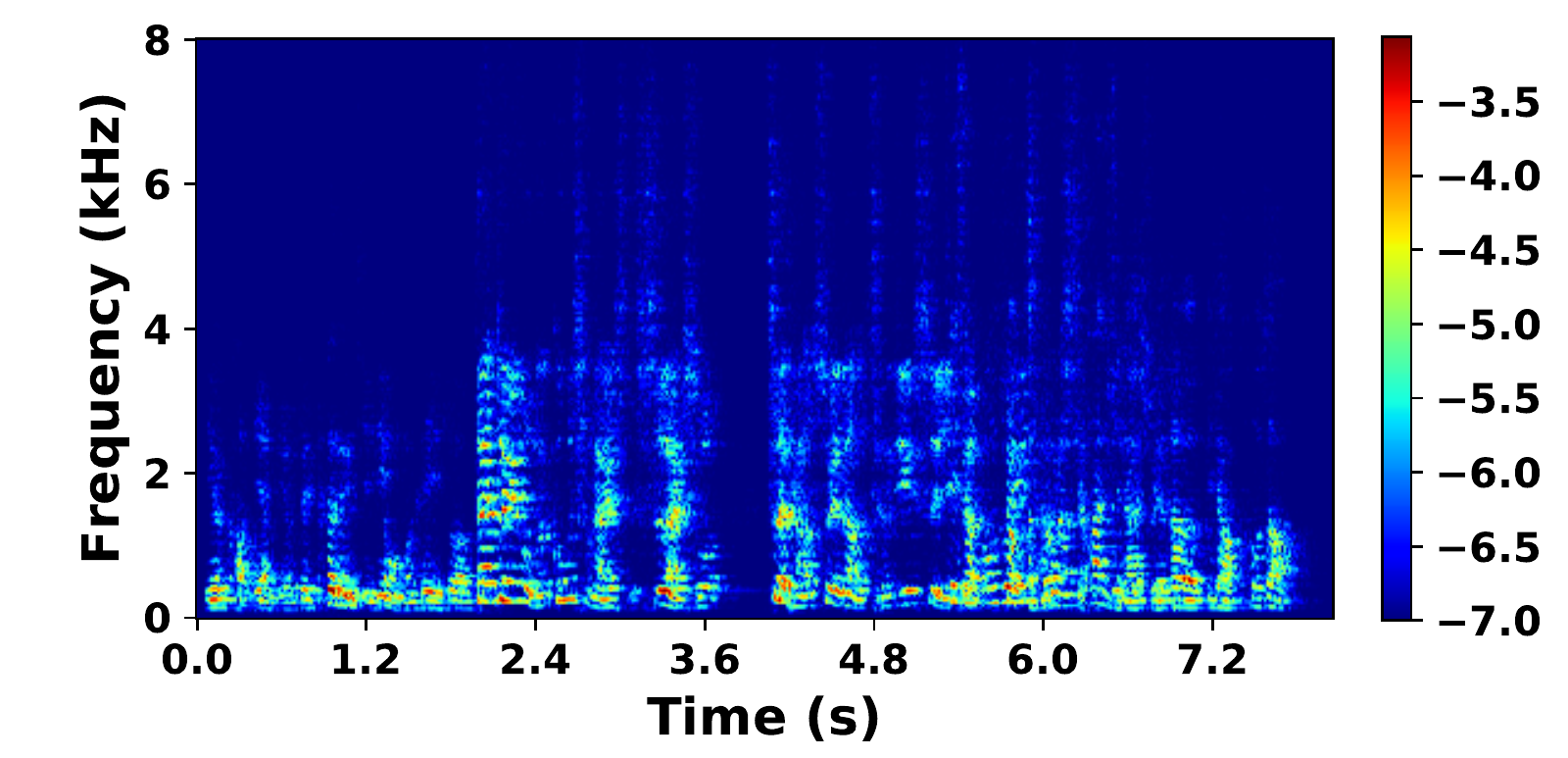}%
		\label{fig:subfig:mix}%
	}%
	
	\subfigure[Anechoic Target Speech]{%
		\includegraphics[width=8cm]{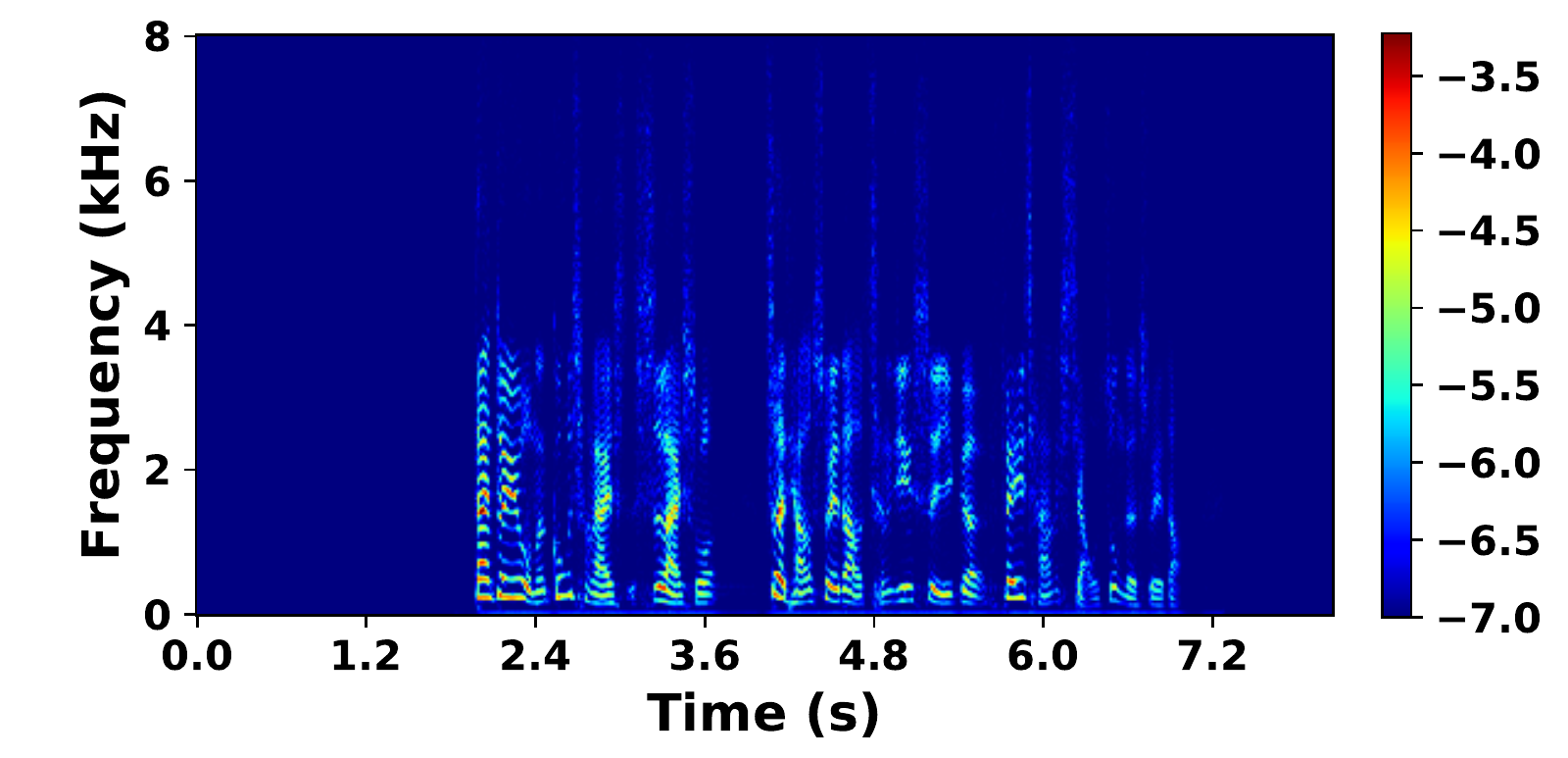}%
		\label{fig:subfig:s}%
	}%
	
	\subfigure[Estimated Reverberant Speech (Resynthesized from the Output of the Separation Module)]{%
		\includegraphics[width=8cm]{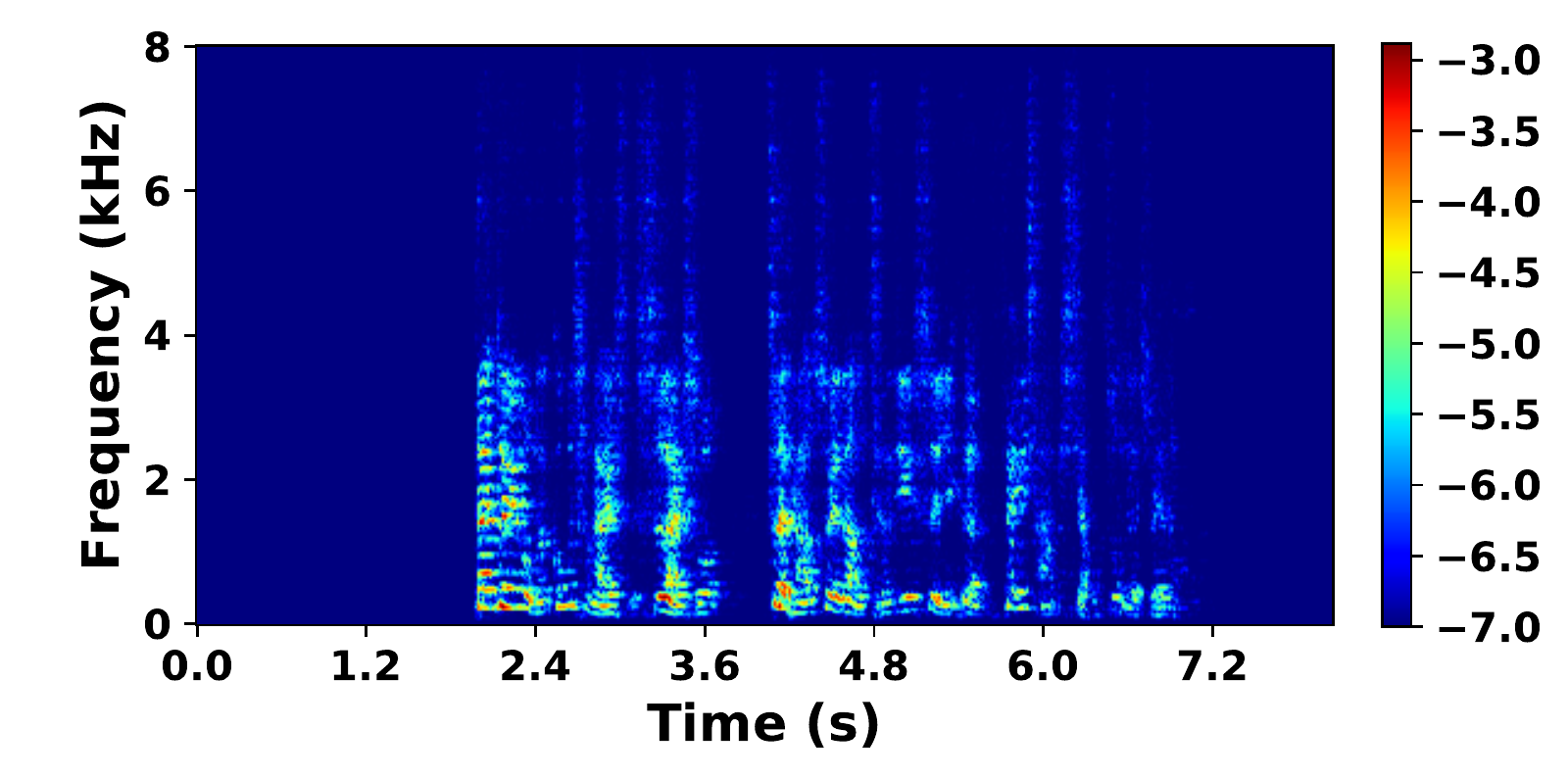}%
		\label{fig:subfig:s_reverb_est}%
	}%

	\subfigure[Estimated Anechoic Speech]{%
		\includegraphics[width=8cm]{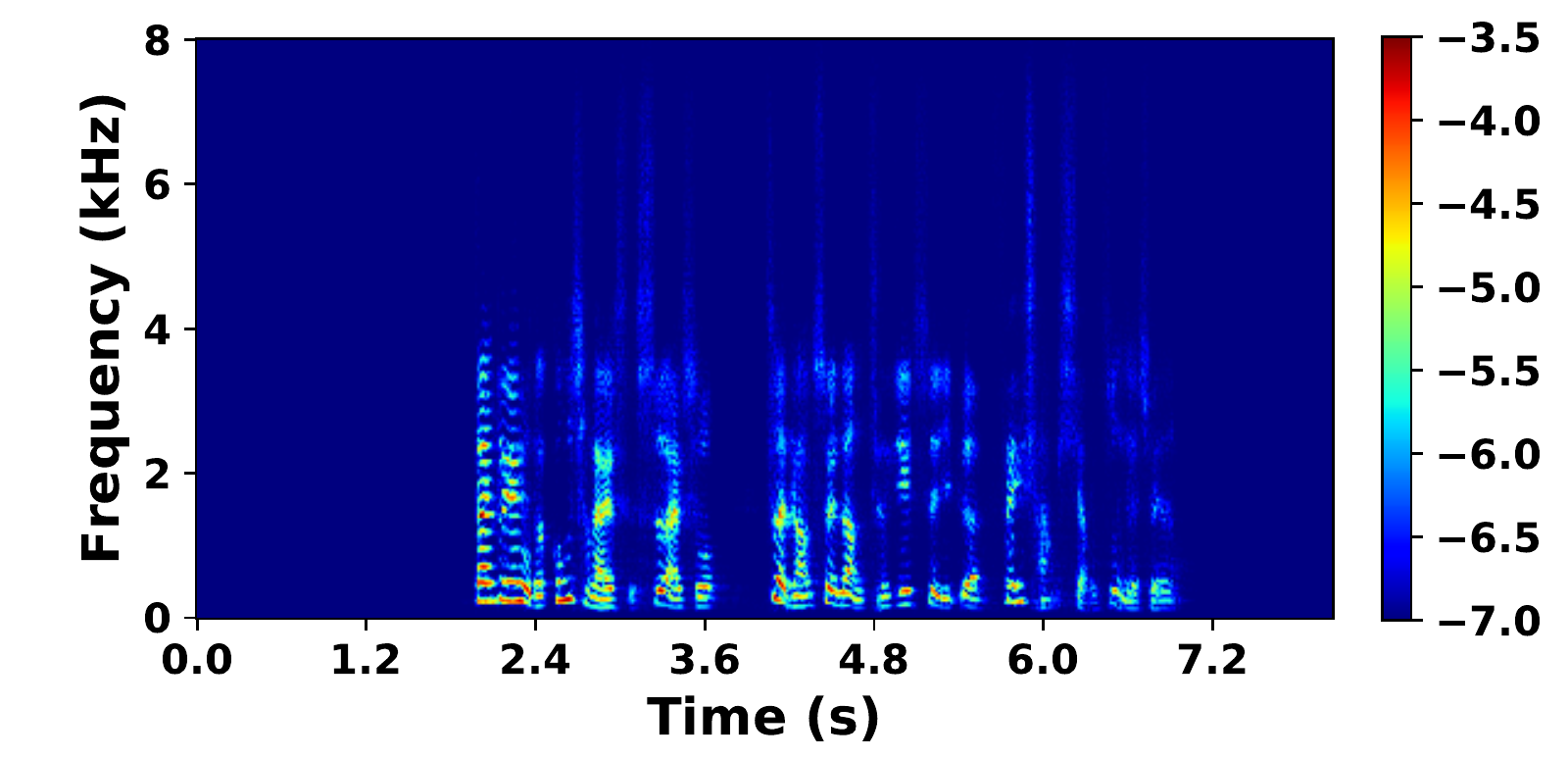}%
		\label{fig:subfig:s_est}%
	}%
	
	\caption{(Color Online). An example of the spectrograms of an unprocessed three-speaker mixture, anechoic target speech (ground-truth) and estimated speech by our proposed two-stage multimodal network ($\lambda$=0.08). The spectral magnitudes are plotted on a log scale.}
	\label{fig:spec}
\end{figure}

We now compare two-stage approaches with IDs 5-7 and 9-16, in which separation is performed in the first stage and dereverberation in the second stage. In approaches 5 and 6, the separation module and the dereverberation module in Fig.~\ref{fig:architecture} are well trained separately, while the whole network is not jointly trained. Specifically, approach 5 trains the BLSTM with an SI-SNR loss $\mathcal{L}_{\text{SI-SNR}}$ in the dereverberation stage, and approach 6 with an MSE loss $\mathcal{L}_{\text{MSE}}$. As shown in Table~\ref{tab:estoi_pesq}, these two approaches produce similar ESTOI, while approach 5 yields a 0.12 PESQ improvement over approach 6. It should be pointed out that, ESTOI is designed to measure the objective intelligibility of speech, and is inappropriate for evaluation of dereverberation. Moreover, both approaches 5 and 6 significantly outperform the one-stage baselines (i.e. 1-4). For example, approach 6 improves ESTOI by 2.33\% and PESQ by 0.19 over approach 4. In addition, approach 6 yields significantly higher ESTOI and PESQ than approach 7, where the BLSTM in the dereverberation module is replaced by a single-channel weighted prediction error (WPE) minimization~\cite{yoshioka2012generalization} method. The WPE method is a representative method for speech dereverberation. Going from approach 6 (without joint training) to approach 9 (with joint training) substantially improves both metrics. Note that an MSE loss $\mathcal{L}_{\text{MSE}}$ (i.e. $\mathcal{L}'_{\text{Multi-Obj}}$ with $\lambda$=0) is used for joint training in approach 9. To further demonstrate the effectiveness of our proposed two-stage approach, we additionally train a network with the same architecture as in Fig.~\ref{fig:architecture}, whereas the whole network is trained from scratch, unlike approach 9 that trains the dilated CNN and the BLSTM in two separate stages prior to joint optimization. This network amounts to a one-stage approach, i.e. approach 8. We can observe that approach 9 dramatically improves ESTOI by 9.16\% and PESQ by 0.36 over approach 8. Further improvements can be achieved by jointly training the dilated CNN and the BLSTM with the multi-objective loss function $\mathcal{L}'_{\text{Multi-Obj}}$ described in Section~\ref{subsec:JT}. We find that $\lambda$=0.08 leads to the best performance in terms of ESTOI and PESQ, which achieves a 21.10\% ESTOI improvement and a 0.79 PESQ improvement over the unprocessed mixtures. For all approaches in Table~\ref{tab:estoi_pesq}, smaller DOA angles correspond to smaller improvements over the unprocessed mixtures in terms of both ESTOI and PESQ, as the angle features become less discriminative and effective when the DOA angle between the target and interfering speech signals decreases.

\begin{table}[t]\scriptsize
	\caption{Relative WER Improvements over the Unprocessed Mixtures.} 
	\newcolumntype{V}{!{\vrule width 0.7pt}}
	\newcolumntype{U}{!{\vrule width 0.7pt}}
	\centering
	%\[
		\begin{tabular}{VcVcV}
			\Xhline{0.7pt}
			\textbf{Approaches} & \textbf{Relative WER Improvements} \\
			\Xhline{0.7pt}
			Dilated CNN (Separation) & 37.13\% \\
			Dilated CNN+WPE & 38.40\% \\
			Dilated CNN+BLSTM ($\lambda$=0.08) (Prop.) & \textbf{46.17\%} \\
			Reverberant Target Speech & 82.06\% \\
			Anechoic Target Speech & 87.95\% \\
			\Xhline{0.7pt}
			
		\end{tabular}
	%\]
	\label{tab:wer_improve}
\end{table}

\begin{table*}[t]\scriptsize
	\caption{ESTOI and PESQ comparisons of different approaches on one-speaker, two-speaker and three-speaker mixtures.} 
	\newcolumntype{V}{!{\vrule width 0.7pt}}
	\newcolumntype{U}{!{\vrule width 0.7pt}}
	\centering
	%\[
	\begin{tabular}{VcVcVcVcVcVcVcVcVcV}
		\Xhline{0.7pt}
		\multirow{2}{*}{\textbf{Approaches}} & \multirow{2}{*}{\textbf{IDs}} & \multirow{2}{*}{\textbf{Loss Functions}} & \multicolumn{3}{cV}{\textbf{ESTOI (in \%)}} &  \multicolumn{3}{cV}{\textbf{PESQ}} \\ \Cline{0.7pt}{4-9} & & & 1-speaker & 2-speaker & 3-speaker & 1-speaker & 2-speaker & 3-speaker \\
		\Xhline{0.7pt}
		Unprocessed & - & - & 66.71 & 46.80 & 40.51 & 2.48 & 1.67 & 1.50 \\
		\Xhline{0.7pt}
		Dilated CNN+BLSTM (w/o JT) & 6 & $\mathcal{L}_{\text{SI-SNR}}$, $\mathcal{L}_{\text{MSE}}$ & 69.58 & 62.86 & 56.59 & 2.56 & 2.26 & 2.11 \\
		Dilated CNN+WPE & 7 & $\mathcal{L}_{\text{SI-SNR}}$~(Separation Stage) & 69.73 & 59.92 & 53.57 & 2.59 & 2.10 & 1.91 \\
		Dilated CNN+BLSTM (One-Stage) & 8 & $\mathcal{L}_{\text{MSE}}$ & 72.31 & 58.04 & 51.30 & 2.64 & 2.06 & 1.86 \\
%		\rowcolor{mygray}
		Dilated CNN+BLSTM (w/ JT) & 13 & $\mathcal{L}_{\text{SI-SNR}}$, $\mathcal{L}_{\text{MSE}}$, $\mathcal{L}'_{\text{Multi-Obj}}$~($\lambda$=0.08) & \textbf{75.09} & \textbf{67.72} & \textbf{61.99} & \textbf{2.77} & \textbf{2.45} & \textbf{2.28} \\
		\Xhline{0.7pt}		
	\end{tabular}
	%\]
	\label{tab:single_speaker}
\end{table*}

Moreover, SI-SNR improvements (SI-SNRi) over the unprocessed mixtures are shown in Fig.~\ref{fig:si-snri}, where the SI-SNRi is calculated as $\text{SI-SNRi} = \text{SI-SNR}_{\text{processed}} - \text{SI-SNR}_{\text{unprocessed}}$. It can be observed that approaches 3 and 5, which train the network to directly maximize SI-SNR, yield better SI-SNRi than approaches 4 and 6. Our proposed approaches (i.e. 10-16) produce better SI-SNRi than approach 7 (Dilated CNN+WPE) and approach 8 (one-stage). When $\lambda$=0.08, our proposed two-stage approach improves SI-SNR by 5.91~dB over the unprocessed mixtures. Further increasing $\lambda$ leads to slightly higher SI-SNR, as a larger value of $\lambda$ leads to more emphasis on the SI-SNR loss during training. Fig.~\ref{fig:spec} shows an example of the spectrograms of an unprocessed mixture, anechoic target speech and estimated speech by our proposed two-stage multimodal network ($\lambda$=0.08). More demos can be found at~\footnote{https://jupiterethan.github.io/av-enh.github.io/}. We additionally evaluate our proposed approach, as well as two baselines, using the web version of Google's Chinese Mandarin speech recognition engine\footnote{https://www.google.com/intl/en/chrome/demos/speech.html}. The relative improvements of word error rates (WERs) over the unprocessed mixtures are presented in Table~\ref{tab:wer_improve}, where the relative WER improvement is calculated as $(\text{WER}_{\text{unprocessed}} - \text{WER}_{\text{processed}}) / \text{WER}_{\text{unprocessed}}$. Note that the unprocessed mixtures produce a WER of 92.90\%. The first baseline in Table~\ref{tab:wer_improve} is a dilated CNN with the same architecture as the separation module in Fig.~\ref{fig:architecture}, which is trained to separate the reverberant target speech from interfering speech and background noise. In other words, dereverberation is not performed in this baseline, which yields a 37.13\% relative WER improvement over the unprocessed mixtures. A slightly larger relative improvement is achieved by additionally using a WPE method for dereverberation. As shown in Table~\ref{tab:wer_improve}, our proposed approach produces a 46.17\% relative WER improvement compared with the unprocessed mixtures, which is significantly better than the two baselines. In addition, anechoic target speech (i.e. ground-truth) produces a 87.95\% relative WER improvement over the unprocessed mixtures, which provides an upper bound for the separation and dereverberation systems.

We further evaluate approaches 6, 7, 8 and 13 on a set of one-speaker mixtures, which are created by mixing target speech and background noise in reverberant environments. Table~\ref{tab:single_speaker} shows the ESTOI and PESQ results on mixtures with different numbers of speakers, where all models are trained on two-speaker and three-speaker mixtures. It can be observed that, approaches 6 and 7 only yield slight improvements on ESTOI and PESQ over the unprocessed mixtures, and further improvements can be obtained by approach 8. Note that approaches 6-8 exhibit a performance trend on one-speaker mixtures opposite to that on two-speaker and three-speaker mixtures. Our proposed approach (i.e. approach 13) produces consistently higher ESTOI and PESQ than the three baselines, which reveals its relatively stronger generalization capability in the single-speaker scenario. In addition, we can see that the problem becomes more difficult with more interfering speakers.

\begin{figure}[t]
	\centering
	
	\subfigure[ESTOI vs. DRR Range]{%
		\includegraphics[width=4.4cm]{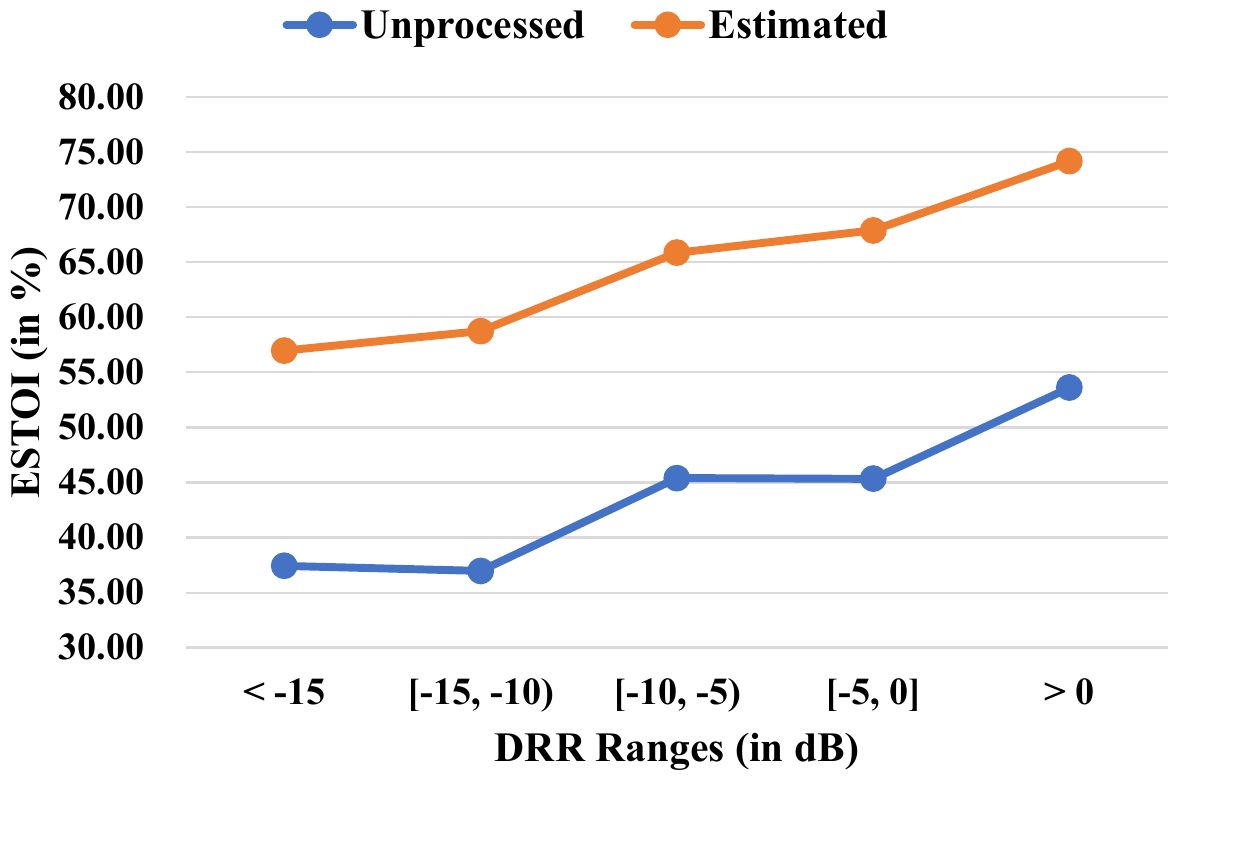}%
		\label{fig:subfig:estoi_drr}%
	}%
	\subfigure[PESQ vs. DRR Range]{%
		\includegraphics[width=4.4cm]{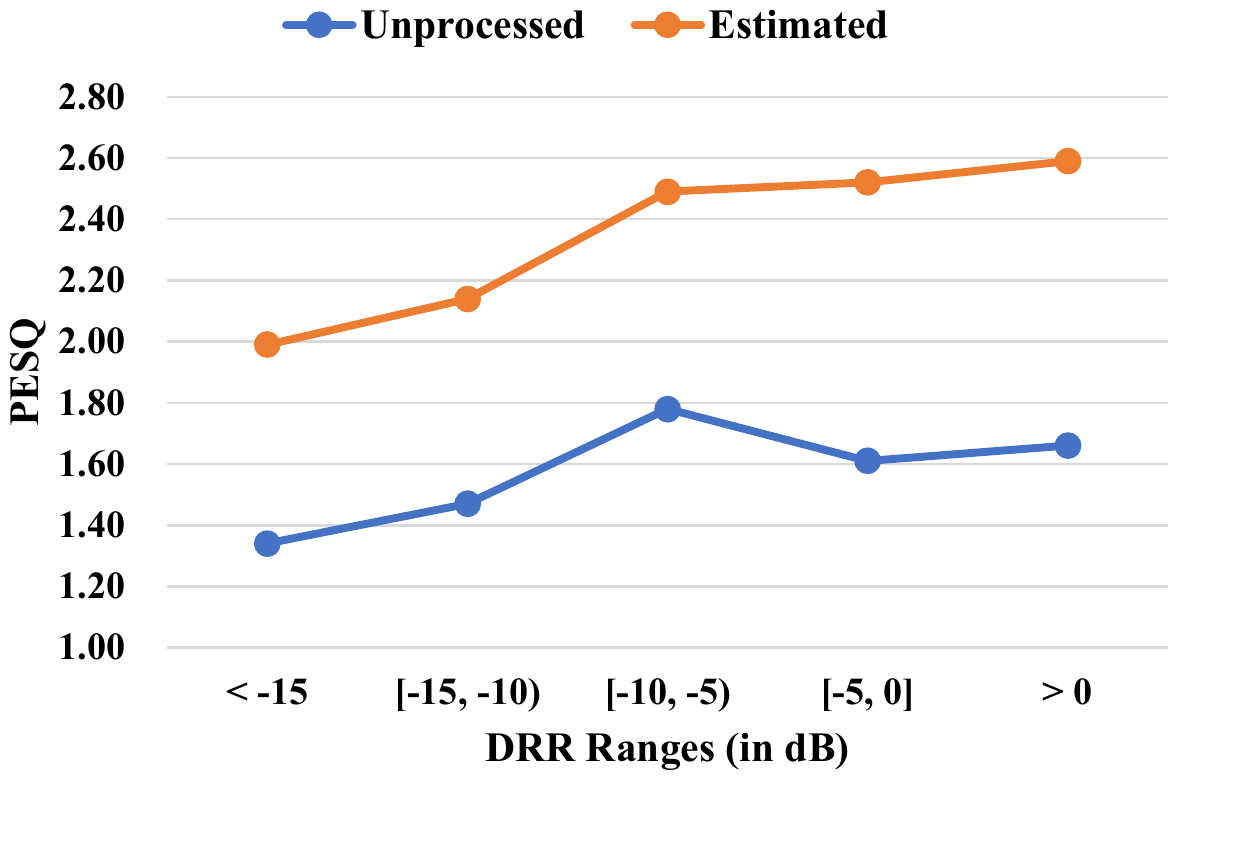}%
		\label{fig:subfig:pesq_drr}%
	}%

	\subfigure[$\Delta$ESTOI vs. DRR Range]{%
		\includegraphics[width=4.4cm]{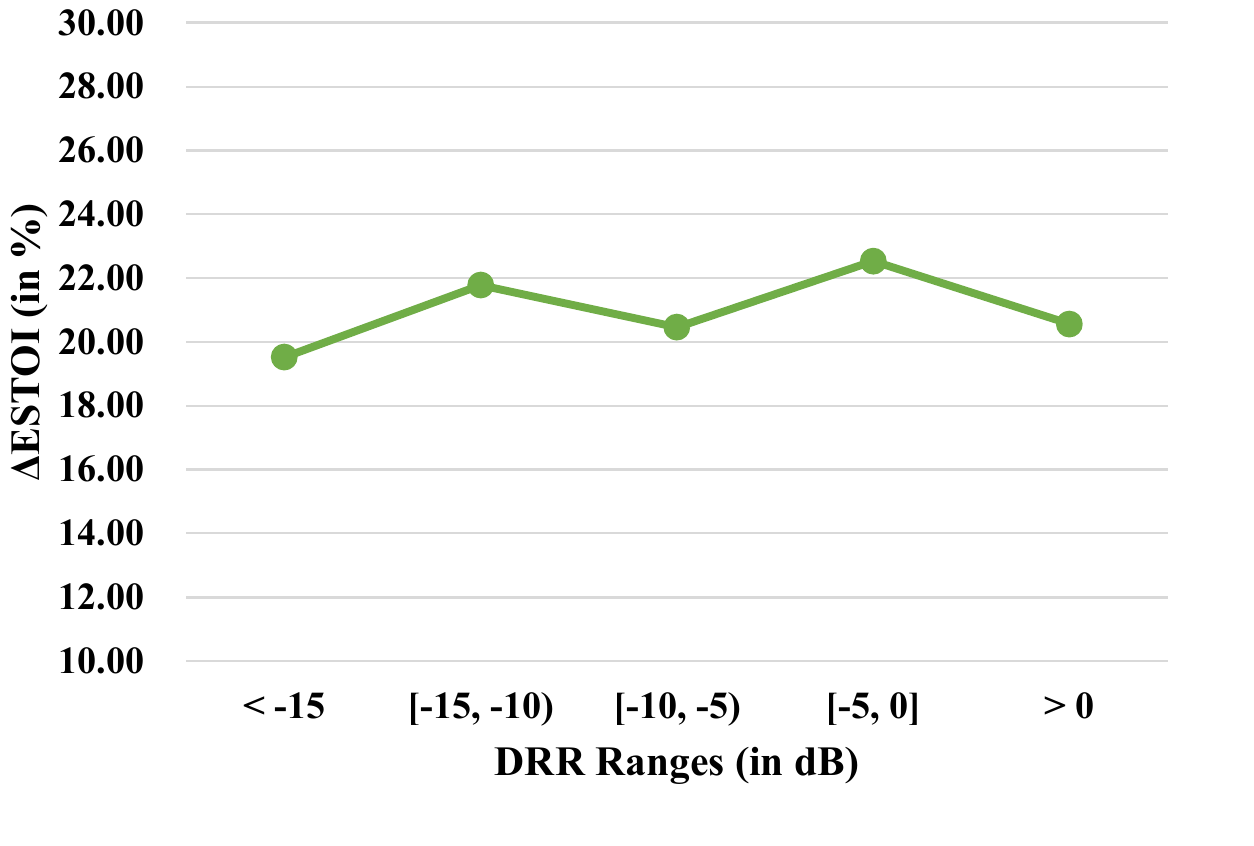}%
		\label{fig:subfig:delta_estoi_drr}%
	}%
	\subfigure[$\Delta$PESQ vs. DRR Range]{%
		\includegraphics[width=4.4cm]{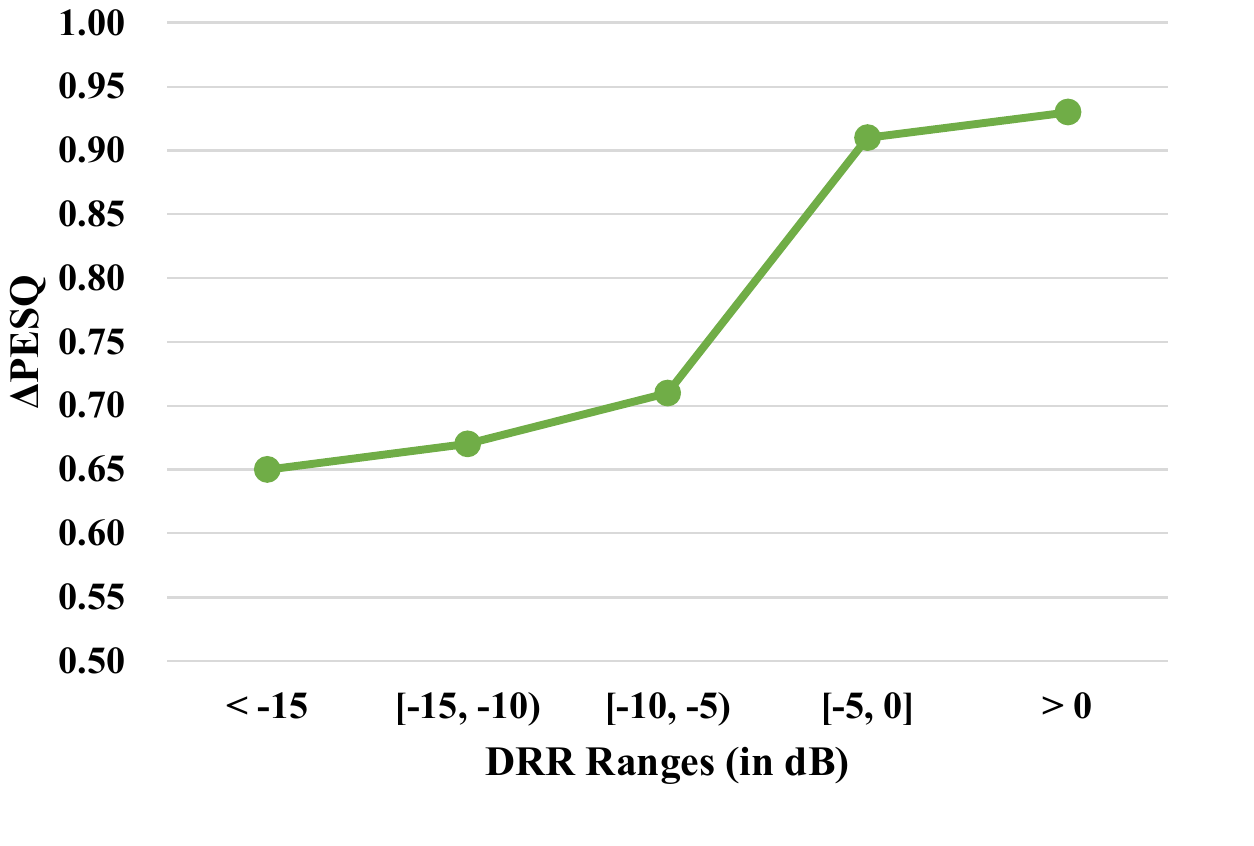}%
		\label{fig:subfig:delta_pesq_drr}%
	}%
		
	\caption{(Color Online). ESTOI and PESQ results for different DRRs. $\Delta$ESTOI and $\Delta$PESQ indicate the ESTOI and PESQ improvements over unprocessed mixtures, respectively.}
	\label{fig:drr}
\end{figure}

In addition, the ESTOI and PESQ results produced by the proposed approach (with ID 13) for different direct-to-reverberant ratio (DRR)~\cite{naylor2010speech} ranges are presented in Figs.~\ref{fig:subfig:estoi_drr} and~\ref{fig:subfig:pesq_drr}. The improvements over unprocessed mixtures are shown in Figs.~\ref{fig:subfig:delta_estoi_drr} and~\ref{fig:subfig:delta_pesq_drr}. The DRR is calculated from the RIR corresponding to the target speaker as follows:
\begin{equation}
\text{DRR} = 10 \log_{10}{\frac{\sum_{k=0}^{k_d}{h_s^2(k)}}{\sum_{k=k_d+1}^{\infty}{h_s^2(k)}}}~\text{dB},
\end{equation}
where $h_s$ represents the RIR corresponding to the target speaker at the first microphone, and $k_d$ is a time index that separates the RIR into
two parts, one for the direct-path propagation and another for reverberation
due to reflected paths. Note that the unprocessed samples with higher DRRs do not always correspond to higher PESQ values due to a variety of other factors such as the randomly selected SNRs and TIRs. It can be observed that $\Delta$ESTOI does not exhibit a clear trend for different DRR ranges, as DRR is mostly related to speech quality rather than objective intelligibility. Higher DRRs correspond to larger $\Delta$PESQ, as shown in Fig.~\ref{fig:subfig:delta_pesq_drr}.

\subsection{Investigation of Separation Performance}
We now investigate the performance of the separation module. The independently trained separation module (prior to joint training) is used for investigation. We use the same dilated CNN with PIT as a baseline model. Note that we do not use the angle features and lip features in the PIT baseline, as these features are speaker-specific or direction-specific and thus inappropriate for the PIT setup. The PIT baseline has two outputs, one for target reverberant speech and another for the residual signal. The residual signal is derived by subtracting target reverberant speech from the first-channel mixture, which contains interfering reverberant speech and background noise. We train the PIT baseline using the SI-SNR loss in Eq.~(\ref{eq:sisnr1}). As shown in Table~\ref{tab:separation}, the proposed separation module significantly outperforms the PIT baseline, which indicates that the visual information as well as the angle features improves the separation performance.

\begin{table}[t]\scriptsize
	\caption{Investigation of Separation Performance.} 
	\newcolumntype{V}{!{\vrule width 0.7pt}}
	\newcolumntype{U}{!{\vrule width 0.7pt}}
	\centering
	%\[
	\begin{tabular}{VcVcVcVcV}
		\Xhline{0.7pt}
		\textbf{Approach} & \textbf{ESTOI (in \%)} & \textbf{PESQ} & \textbf{SI-SNRi (in dB)} \\
		\Xhline{0.7pt}
		Unprocessed & 59.02 & 1.93 & - \\
		\Xhline{0.7pt}
		PIT & 64.53 & 2.17 & 4.68 \\
		Prop. & \textbf{69.91} & \textbf{2.47} & \textbf{10.04} \\	
		\Xhline{0.7pt}
		
	\end{tabular}
	%\]
	\label{tab:separation}
\end{table}

\begin{table*}[t]\scriptsize
	\caption{Investigation of the impact of visual information on separation and dereverberation in the jointly trained model.} 
	\newcolumntype{V}{!{\vrule width 0.7pt}}
	\newcolumntype{U}{!{\vrule width 0.7pt}}
	\centering
	%\[
	\begin{tabular}{VcVcVcVcVcVcV}
		\Xhline{0.7pt}
		\textbf{Approaches} & \textbf{Use Lip for Sep.} & \textbf{Use Lip for Dereverb.} &\textbf{ESTOI (in \%)} & \textbf{PESQ} & \textbf{SI-SNRi (in dB)} \\
		\Xhline{0.7pt}
		Unprocessed & - & - & 44.21 & 1.60 & - \\
		\Xhline{0.7pt}
		Dilated CNN+BLSTM ($\lambda$=0.08) (Prop.) & \cmark & \xmark & 65.31 & \textbf{2.39} & 5.91 \\
		Dilated CNN+BLSTM ($\lambda$=0.08) & \xmark & \xmark & 61.47 & 2.14 & 5.14 \\
		Dilated CNN+BLSTM ($\lambda$=0.08) & \cmark & \cmark & \textbf{65.77} & \textbf{2.39} & \textbf{6.03} \\
		\Xhline{0.7pt}
		
	\end{tabular}
	%\]
	\label{tab:visual_dereverb}
\end{table*}

\subsection{Impact of Visual Features on Separation and Dereverberation}
The impact of visual information on separation and dereverberation is further investigated in this section. We first compare the proposed method with a new baseline model, which does not use any visual inputs throughout the network. The baseline can be easily derived by removing the visual submodule from the multimodal network shown in Fig.~\ref{fig:architecture}. Table~\ref{tab:visual_dereverb} lists the ESTOI, PESQ and SI-SNRi results. We can observe that the removal of visual information significantly degrades the performance in the three metrics, which suggests the effectiveness of visual inputs. To investigate the impact of visual features on dereverberation, we additionally train a baseline model for comparison, which uses visual embeddings in both the separation module and the dereverberation module. The visual embeddings used in both modules are learned by the same visual submodule. Specifically, we concatenate the output spectra of the separation module with the visual embeddings, which are then fed into the BLSTM for dereverberation. Note that the separated spectra and the visual embeddings are passed through a layer normalization layer prior to concatenation. As shown in Table~\ref{tab:visual_dereverb}, the inclusion of visual features for dereverberation yields very slight improvements in ESTOI and SI-SNRi, and no improvement in PESQ, over the proposed method. An interpretation is that the dereverberation module implicitly benefits from the visual features in the separation module due to joint training. Hence, it is unnecessary to explicitly use visual features in the dereverberation module.

\subsection{Robustness Against Missing Visual Frames}
In real applications, the lip images of speakers are not always captured, particularly when they do not face towards the camera temporarily. These lip images are recognized as missing visual inputs. In this case, we compensate for the missing lip images in the following way. For a missing frame of target or interfering speakers' lips, it is filled with the latest existing frame. To investigate the robustness of our proposed method against missing lip information, we randomly discard frames for each speaker and apply the compensation method during inference. We investigate three scenarios: (1) lip information of only the interfering speaker(s) is partially lost; (2) lip information of only the target speaker is partially lost; (3) lip information of all speakers is partially lost. 

Table~\ref{tab:lost_lip} shows that the proposed method is robust against missing lip information for a frame loss rate of 40\%, even when all speakers' lip information is incomplete. For a visual frame loss rate of 80\%, the performance almost does not degrade if only interfering speakers' lip information is partially missing. However, both ESTOI and PESQ significantly decrease when 80\% of target speaker's lip images are missing. This decrease becomes more moderate if the interfering speakers' lip information is complete, which indicates that the presence of interfering speakers' lip information improves the robustness against lost visual frames. Moreover, we would like to point out that our system still works even if 80\% of all speakers' lip frames are missing. This is likely due to the use of angle features (see Section~\ref{subsec:af}), which provide useful directional cues for the target speaker.

\begin{table}[t]\scriptsize
	\caption{Investigation of the robustness against incomplete lip information.} 
	\newcolumntype{V}{!{\vrule width 0.7pt}}
	\newcolumntype{U}{!{\vrule width 0.7pt}}
	\centering
	%\[
	\begin{tabular}{VcVcVcVcV}
		\Xhline{0.7pt}
		\textbf{Lip Image Loss Rate} & \textbf{Which Speaker Is Lost} & \textbf{ESTOI (in \%)} & \textbf{PESQ} \\
		\Xhline{0.7pt}
		0\% & None & 65.31 & \textbf{2.39} \\
		\Xhline{0.7pt}
%		40\% (w/o Comp.) & Interference & 64.62 & 2.34 \\
%		40\% (w/o Comp.) & Target & 36.92 & 1.14 \\
		40\% & Interference & \textbf{65.32} & 2.38 \\
		80\% & Interference & 65.31 & 2.38 \\
		40\% & Target & \textbf{65.32} & 2.38 \\
		80\% & Target & 63.34 & 2.20 \\
		40\% & All & 65.21 & 2.37 \\
		80\% & All & 62.87 & 2.18 \\
		\Xhline{0.7pt}
		
	\end{tabular}
	%\]
	\label{tab:lost_lip}
\end{table}

\subsection{Robustness Against Reduced Lip Image Resolution}
The resolution of lip images can be low due to the poor quality of the camera in practice. To investigate the robustness of the proposed method against lower image resolution, we reduce all speakers' lip image resolution for the testing samples. Specifically, we first downsample the lip images from 112$\times$112 to 64$\times$64, and then upsample them back to 112$\times$112. Such an operation reduces the image resolution. As shown in Table~\ref{tab:lower_res}, the proposed model is robust against reduced image resolution.

\begin{table}[t]\scriptsize
	\caption{Investigation of the robustness against reduced image resolution.} 
	\newcolumntype{V}{!{\vrule width 0.7pt}}
	\newcolumntype{U}{!{\vrule width 0.7pt}}
	\centering
	%\[
	\begin{tabular}{VcVcVcVcV}
		\Xhline{0.7pt}
		\textbf{Resolution} & \textbf{ESTOI (in \%)} & \textbf{PESQ} & \textbf{SI-SNRi (in dB)} \\
		\Xhline{0.7pt}
		Original & \textbf{65.31} & \textbf{2.39} & \textbf{5.91} \\
		Reduced & 65.30 & 2.38 & 5.89 \\
		\Xhline{0.7pt}
		
	\end{tabular}
	%\]
	\label{tab:lower_res}
\end{table}

\section{Concluding Remarks}
\label{sec:conclusion}
In this study, we have proposed a two-stage multimodal network for audio-visual separation and dereverberation in noisy and reverberant environments, motivated by the fact that additive interference (e.g. interfering speech and background noise) and convolutive interference (e.g. room reverberation) distort target speech in intrinsically different ways. A dilated CNN based separation module, which takes both audio and visual inputs, is employed to separate reverberant target speech from interfering speech and background noise. The output of the separation module is subsequently passed through a BLSTM based dereverberation module. The two modules are first trained separately and then trained jointly to optimize a new multi-objective loss function, which combines a time domain loss and a T-F domain loss. Systematic evaluations show that our proposed two-stage multimodal network consistently outperforms several one-stage and two-stage baselines in terms of both objective intelligibility and perceptual quality. We find that the proposed network substantially improves ESTOI and PESQ over the unprocessed mixtures. In addition, our network architecture can accept visual streams from an arbitrary number of interfering speakers, which is more advantageous than the multimodal networks that do not allow the number of speakers to change from training to testing.

It should be noted that the proposed model is a noncausal system, which utilizes a large amount of future information for estimation. Such a model is inapplicable to real-time processing, which is highly demanded by many real-world applications. We have preliminarily investigated some causal and partially causal models for real-time processing, with no or low latency. For future work, we would devote more efforts to the design of new multimodal network architectures for real-time speech separation and dereverberation in far-field scenarios.

% or


%\appendices

%\section{}

%\section*{Acknowledgment}


\ifCLASSOPTIONcaptionsoff
  \newpage
\fi

\bibliographystyle{abbrv}
\bibliography{./bare_jrnl}

%\begin{IEEEbiography}{Ke Tan}
%\end{IEEEbiography}
%\begin{IEEEbiographynophoto}{Yong Xu}

%\newpage


%\vfill


\end{document}